%% file: tudor_maveric_south_post_acceptance.tex
\documentclass[fleqn,usenatbib]{mnras}
\usepackage{newtxtext,newtxmath}
\usepackage{multirow}

\usepackage[T1]{fontenc}
\usepackage{ae,aecompl}

\usepackage{graphicx}	
\usepackage{amsmath}	
\usepackage[compatibility=false]{caption}
\usepackage{subcaption}
\usepackage{array}
\usepackage{lscape}
\usepackage{siunitx} 

\DeclareCaptionLabelFormat{continued}{Figure#1~#2 (Continued)}


\title[Radio survey of southern globular clusters]{The MAVERIC survey: A catalogue of radio sources in southern globular clusters from the Australia Telescope Compact Array}

\author[V. Tudor et al.]{\parbox[T]{18cm}{
Vlad Tudor$^1$, 
James C.~A.~Miller-Jones$^1$\thanks{E-mail: james.miller-jones@curtin.edu.au}, 
Jay Strader$^2$, 
Arash Bahramian$^1$, 
Laura Shishkovsky$^2$, 
Richard M.~Plotkin$^3$, 
Laura Chomiuk$^2$, 
Craig O.~Heinke$^4$, 
Thomas J.~Maccarone$^5$, 
Gregory R.~Sivakoff$^4$, 
Evangelia Tremou$^6$, 
Gemma E.~Anderson$^1$, 
Thomas D.~Russell$^7$, 
Anastasios K.~Tzioumis$^8$}
\vspace{0.2cm}\\
$^{1}$International Centre for Radio Astronomy Research -- Curtin University, GPO Box U1987, Perth, WA 6845, Australia\\
$^{2}$Center for Data Intensive and Time Domain Astronomy, Department of Physics and Astronomy, Michigan State University, East Lansing, MI 48824, USA\\
$^{3}$Department of Physics, University of Nevada, Reno, NV 89557, USA\\
$^{4}$Department of Physics, University of Alberta, CCIS 4-183, Edmonton, AB T6G 2E1, Canada\\
$^{5}$Department of Physics, Box 41051, Science Building, Texas Tech University, Lubbock, TX 79409, USA\\
$^{6}$LESIA, Observatoire de Paris, CNRS, PSL Research University, Sorbonne Universite, Universite de Paris, Meudon, France\\
$^{7}$INAF, Istituto di Astrofisica Spaziale e Fisica Cosmica, Via U. La Malfa 153, I-90146 Palermo, Italy\\
$^{8}$Australia Telescope National Facility, CSIRO, PO Box 76, Epping, New South Wales 1710, Australia\\
}

\date{Accepted XXX. Received YYY; in original form ZZZ}

\pubyear{2022}

\hypersetup{draft}
\begin{document}
\label{firstpage}
\pagerange{\pageref{firstpage}--\pageref{lastpage}}
\maketitle

\graphicspath{{./figures/}}

\begin{abstract}
Radio continuum observations offer a new window on compact objects in globular clusters compared to typical X-ray or optical studies. As part of the MAVERIC survey, we have used the Australia Telescope Compact Array to carry out a deep (median central noise level $\approx 4$\,\si{\micro}Jy\,beam$^{-1}$) radio continuum survey of 26 southern globular clusters at central frequencies of 5.5 and 9.0 GHz. This paper presents a catalogue of 1285 radio continuum sources in the fields of these 26 clusters. Considering the surface density of background sources, we find significant evidence for a population of radio sources in seven of the 26 clusters, and also identify at least 11 previously known compact objects (6 pulsars and 5 X-ray binaries). While the overall density of radio continuum sources with 7.25-GHz flux densities $\gtrsim 20$\,\si{\micro}Jy in typical globular clusters is relatively low, the survey has already led to the discovery of several exciting compact binaries, including a candidate ultracompact black hole X-ray binary in 47 Tuc.
Many of the unclassified radio sources near the centres of the clusters are likely to be true cluster sources, and multi-wavelength follow-up will be necessary to classify these objects and better understand the demographics of accreting compact binaries in globular clusters.

\end{abstract}



\section{Introduction}

Globular clusters are abundant in X-ray sources, as a result of dynamical encounters between stars and compact objects in high density environments \citep{1975ApJ...199L..93C}. In the Galactic field
a substantial fraction of X-ray transients host stellar-mass black holes (e.g., \citealt{2014MNRAS.437.3087K}). However, virtually all X-ray transients in Galactic globular clusters have so far been found to be neutron stars based on the detection of X-ray thermonuclear bursts or pulsations \citep[see, e.g.,][]{2014ApJ...780..127B}, raising the question of whether the black hole populations residing in globular clusters are different than those in the Galactic field. Initially, it was proposed that few black holes would reside in present-day clusters, as most of them would have been ejected due to mutual interactions \citep{1993Natur.364..421K, 1993Natur.364..423S}. This hypothesis was questioned with the discovery of variable ultraluminous X-ray sources (ULXs) associated with extragalactic globular clusters, many of which are likely associated with stellar-mass black holes \citep{2007Natur.445..183M, 2010ApJ...721..323S, 2010ApJ...725.1805B, 2010ApJ...712L...1I, 2011MNRAS.410.1655M}, unlike the case for ULXs in the field of star-forming galaxies, which may largely be neutron stars \citep[e.g.,][]{2014Natur.514..202B}.

Seven black hole candidates have also been discovered in quiescence in Galactic globular clusters through deep radio observations \citep{2012Natur.490...71S, 2013ApJ...777...69C, 2015MNRAS.453.3918M, 2018ApJ...855...55S, Urquhart20, Zhao20}, although they remain to be dynamically confirmed. 

A rich collection of recent theoretical work now suggests that a significant fraction of black holes may be retained in clusters to the present day \citep{2013MNRAS.430L..30S, 2015ApJ...800....9M, 2016MNRAS.462.2333P, 2016MNRAS.463.2109R,2020ApJS..247...48K,2020ApJ...898..162W}. In addition, the first dynamically-confirmed stellar-mass black holes in globular clusters have recently been found in NGC\,3201 (although they are not transferring mass; \citealt{2018MNRAS.475L..15G,2019A&A...632A...3G,Paduano22}), demonstrating that black holes can indeed be retained in globular clusters.

All of the recent stellar-mass black hole candidates with quiescent accretion were selected via their radio continuum emission, thought to come from compact, partially self-absorbed jets. If the radio-selected black hole candidates are indeed black holes, the lack of discoveries of black hole transients in globular clusters suggests they may have different properties than field black holes. One reason for this may lie in their formation history. Black holes in dynamically-formed binaries may have short orbital periods, translating to faint quiescent and peak outburst luminosities, and short duty cycles \citep{2004ApJ...601L.171K, 2014MNRAS.437.3087K}. Transient outbursts from black holes may therefore be too rare and too faint to be picked up by all-sky X-ray monitors, which is the source of discovery of most dynamically-confirmed Galactic black holes ($\approx$20--25; \citealp{2016A&A...587A..61C, 2016ApJS..222...15T}). Hence, the current sample is biased towards those sources which have bright (detectable by all-sky X-ray monitors, with peak luminosities exceeding $10^{36}$\,erg\,s$^{-1}$) and frequent (every $\sim$50 years or less) outbursts.

Other observing strategies aimed at discovering different populations of black holes have been developed. These include multi-epoch optical spectroscopic surveys like the Apache Point Observatory Galactic Evolution Experiment \citep[APOGEE;][]{Majewski2017}, which has been used to discover a non-accreting black hole via orbitally-induced radial velocity variations (\citealt{2019Sci...366..637T}; see also \citealt{2021MNRAS.504.2577J}); emission-line surveys of the Galactic field (e.g., \citealt{casares18}); the Galactic Bulge Survey, an X-ray and optical survey of the Galactic Centre  \citep{2011ApJS..194...18J}; and the {\it Gaia} mission, an all-sky astrometric survey \citep{2001A&A...369..339P, 2014arXiv1407.6163B, 2017MNRAS.470.2611M, 2017ApJ...850L..13B}. A radio survey is another approach to finding black holes, as originally proposed by \citet{2005MNRAS.360L..30M}. As these surveys use different search strategies, they suffer from different biases, and are thus complementary to each other.

Identifying black holes in globular clusters from radio surveys requires us to understand the populations of other compact radio sources. Within globular clusters, well-known radio emitters are pulsars and X-ray binaries. In deep surveys, fainter sources such as cataclysmic variables or magnetically-active stars could also become detectable. Furthermore, we need to account for the populations of background sources (primarily active galactic nuclei), as well as the unrelated Galactic radio sources that could be present, especially at low Galactic latitude.

Over 230 pulsars (most of which are millisecond pulsars, MSPs, with spin periods below 10\,ms) have so far been discovered in Milky Way globular clusters using radio timing searches \citep{2005ASPC..328..147C, 2008IAUS..246..291R}\footnote{\url{http://www.naic.edu/~pfreire/GCpsr.html}}. The two globular clusters richest in MSPs are Terzan\,5 \citep[with 39 known MSPs;][]{2017ApJ...845..148P} and 47\,Tuc \citep[with 27 known MSPs;][]{2017MNRAS.471..857F}. It is estimated that a few hundred to a few thousand potentially detectable MSPs are present in the Galactic globular cluster system \citep{2005ApJ...625..796H, 2010A&A...524A..75A}.

In the case of X-ray binaries, neutron star radio jets are typically a factor of $\sim$20 fainter than black holes at a similar X-ray luminosity above $10^{36}$\,erg\,s$^{-1}$ \citep{2006MNRAS.366...79M, 2018MNRAS.478L.132G}. However, it is clear that neutron stars show a broader range of radio properties than do black holes (e.g., \citealt{2017MNRAS.470..324T, 2018ApJ...869L..16R, 2020MNRAS.492.1091G, vandenEijnden21, Panurach21}). In addition, the unusual subset of accreting neutron stars known as transitional millisecond pulsars (with $L_X \sim 10^{32}$--$10^{34}$ erg s$^{-1}$) show higher radio luminosities in their accreting states than expected based on the mean relation for other accreting neutron stars, and some are as radio-luminous as black holes
\citep{2011MNRAS.415..235H,2015ApJ...809...13D,2018ApJ...856...54B,Jaodand19,2020ApJ...895...89L,CotiZelati2021}. Overall, the distinction in radio luminosity between accreting black holes and neutron stars has been blurred.

While cataclysmic variables are known to be radio emitters in outburst or persistent states \citep{2007ApJ...660..662M, 2008Sci...320.1318K, 2011MNRAS.418L.129K, 2015MNRAS.451.3801C, 2016MNRAS.463.2229C, 2017AJ....154..252B}, they reach typical peak radio luminosities of $L_{\rm R} \approx 10^{26}$\,erg\,s$^{-1}$. A few unusual systems have been observed at higher luminosities, up to $L_{\rm R} \approx 10^{27}$\,erg\,s$^{-1}$: the white dwarf propeller systems AE Aqr \citep{1993ApJ...406..692A} and LAMOST J024048.51+195226.9 \citep{2021MNRAS.503.3692P}, the white dwarf pulsar AR Sco \citep{2016Natur.537..374M}, and perhaps two intermediate polars \citep{2017AJ....154..252B}.

Magnetically-active stars can also reach radio luminosities of up to $L_{\rm R} \approx 10^{28}$\,erg\,s$^{-1}$ in quiescence, in the case of RS CVn stars \citep{1994A&A...285..621B}. These objects are a class of close binaries consisting of at least one cool giant or sub-giant star \citep{2005LRSP....2....8B} which may be common in globular clusters \citep{2002ASPC..265..277C}. However, above fluxes corresponding to $L_{\rm R} \sim 10^{28}$\,erg\,s$^{-1}$, we expect the detectable radio sources to be dominated by pulsars and X-ray binaries associated with the cluster, as well as background active galactic nuclei. We must also be attuned to the possibility of yet-undiscovered populations of radio sources.

Different classes of radio sources can be roughly classified based on their spectral indices $\alpha$ ($S_\nu \propto \nu^\alpha$, where $S_\nu$ is the flux density at frequency $\nu$). In the GHz range, pulsars have steep spectra ($\alpha = -1.4 \pm 1.0$; \citealp{2000A&AS..147..195M, 2013MNRAS.431.1352B}). Therefore, in radio continuum images, pulsars can most easily be identified as point sources at lower frequencies (1.4\,GHz or lower; \citealp{1990ApJ...363L...5K, 1998ApJS..119...75K, 1999A&AS..136..571H, 2000ApJ...536..865F, 2000MNRAS.311..841M}), but in some cases can also be observed in sufficiently sensitive higher frequency images (5\,GHz; \citealp{1997A&A...318L..63K}). These steep radio spectra can be exploited to find pulsars missed in timing searches due to high or variable scattering along the line of sight \citep{2000A&AS..143..303D, 2018MNRAS.474.5008D}, due to eclipses (e.g., \citealt{Frail18,Urquhart20}) or in close binary systems that have high accelerations. In the hard and quiescent states, X-ray binary jets have flat or slightly inverted radio spectra ($\alpha \approx 0$; \citealp{1979ApJ...232...34B}; \citealp{2001MNRAS.322...31F}). Extragalactic sources that will contaminate radio surveys of globular clusters have a median spectral index $\alpha \sim$ --0.4 to --0.6 at 5\,GHz but with a broad distribution ($-1.5 \gtrsim \alpha \gtrsim +1.0$) that substantially overlaps that expected for the low-mass X-ray binaries and pulsars in globular clusters \citep{1991AJ....102.1258F, 2012MNRAS.426.2342H}. 

Within globular clusters, black holes, neutron stars, and heavy white dwarfs sink towards cluster centres owing to mass segregation, and exotic binaries preferentially form in the densest parts of their cores \citep{2017MNRAS.464.2511H, 2018MNRAS.475.4841R}. Pulsars are known to be more centrally concentrated than the surrounding stars; most are found well within the half-light radius \citep{2009ApJ...690.1370M, 2005ASPC..328..147C}, but some may also be found out to several half-light radii (e.g., \citealp{2002ApJ...570L..89D}), likely due to past interactions. Similarly, black holes are expected to sink towards cluster centres. However, depending on their population within a cluster, black hole heating can sustain large cluster cores, and they can mix with cluster stars out to large radii \citep{2008MNRAS.386...65M, 2015ApJ...800....9M, 2016MNRAS.463.2109R, 2017ApJ...834...68C, 2018MNRAS.477.1853G}. These papers point to an overall theoretical expectation that massive clusters with large cores might be more likely to contain the largest populations of black holes.

In this paper, we catalogue radio continuum sources in deep observations of 26 southern Galactic globular clusters, taken with the Australia Telescope Compact Array (ATCA).
We describe the observations in Section~\ref{sec:obs}. We detail our finding methodology and the source catalogues in Section~\ref{sec:method}. In 
Section~\ref{sec:result} we analyse which of these sources are likely to truly be associated with our globular clusters, and present a discussion of these results in Section~\ref{sec:discuss}.

\section{Observations}
\label{sec:obs}

\subsection{Sample selection}
\label{sec:selection}

The MAVERIC (Milky-way ATCA/VLA Exploration of Radio-sources in Clusters) survey was designed as a volume limited ($D \lesssim 9$\,kpc) sample of  massive clusters ($M \gtrsim 10^5\,M_\odot$), with those limits chosen for sensitivity to quiescent radio emission from accreting black holes, and a restriction to the more massive clusters most likely to host them. We also included a few other more distant, massive globular clusters which may contain intermediate-mass black holes, and others with high interaction rates or bright quiescent X-ray sources.

\begin{figure}
	\includegraphics[width=\columnwidth]{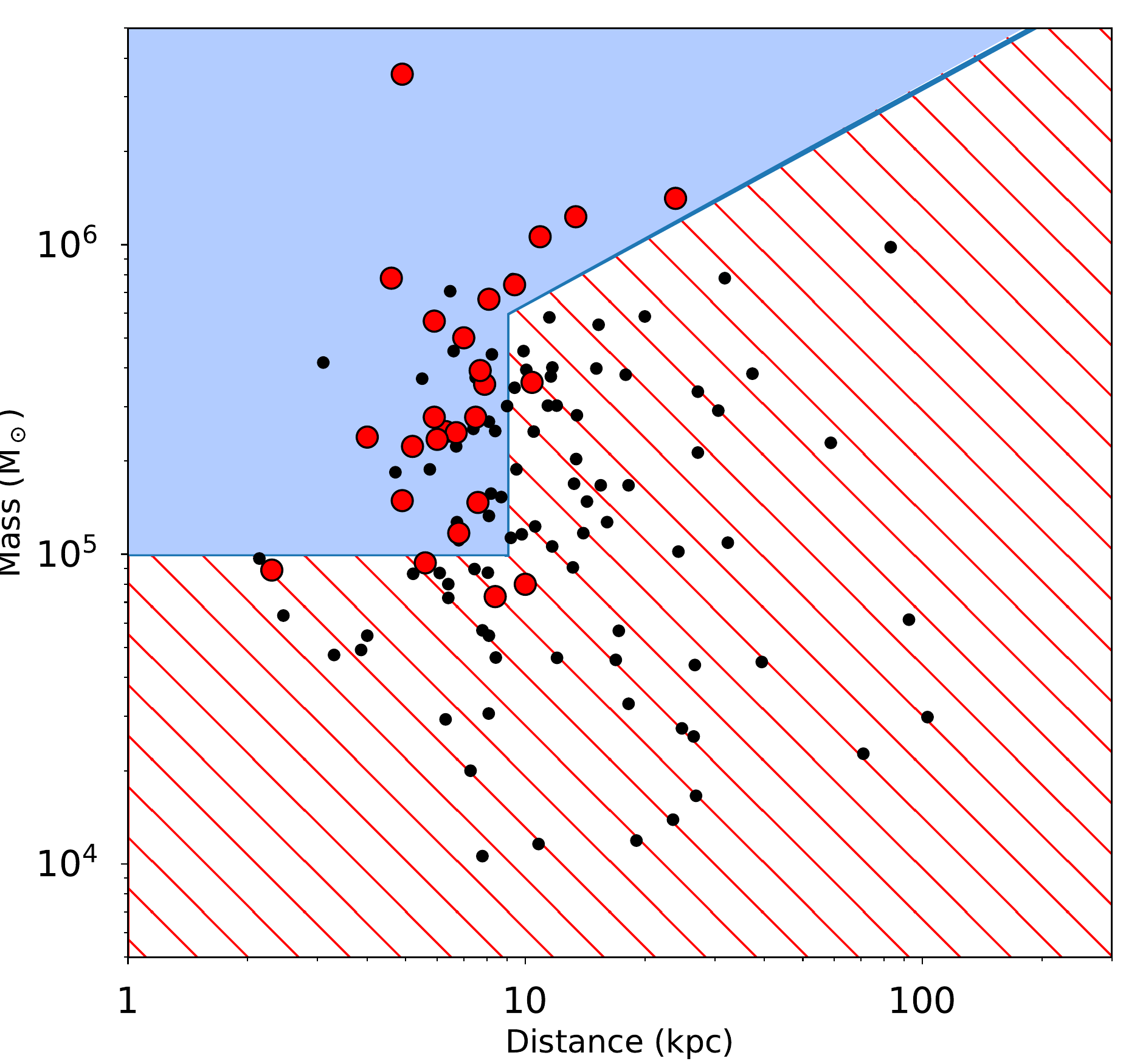}
	\caption{Our globular cluster sample was selected based on mass ($M \gtrsim 10^5\,M_\odot$) and distance ($D \lesssim 9$\,kpc). We also include a few more distant clusters with masses $M \gtrsim 10^6\,M_\odot$, which may contain intermediate-mass black holes, and a few lower-mass clusters with bright X-ray sources. These selection criteria are displayed as a blue region (included sample) and red-hashed region (excluded sample). All globular clusters from the \citet{1996AJ....112.1487H} catalogue are shown in small black circles, and the 26  clusters surveyed with ATCA are shown in large red circles. Those in the blue region that were not imaged with ATCA were located further north ($\delta > -25^{\circ}$) and were instead observed with the VLA \citep{Shishkovsky20}.}
	\label{fig:sample}
\end{figure}

\begin{figure}
	\includegraphics[width=\columnwidth]{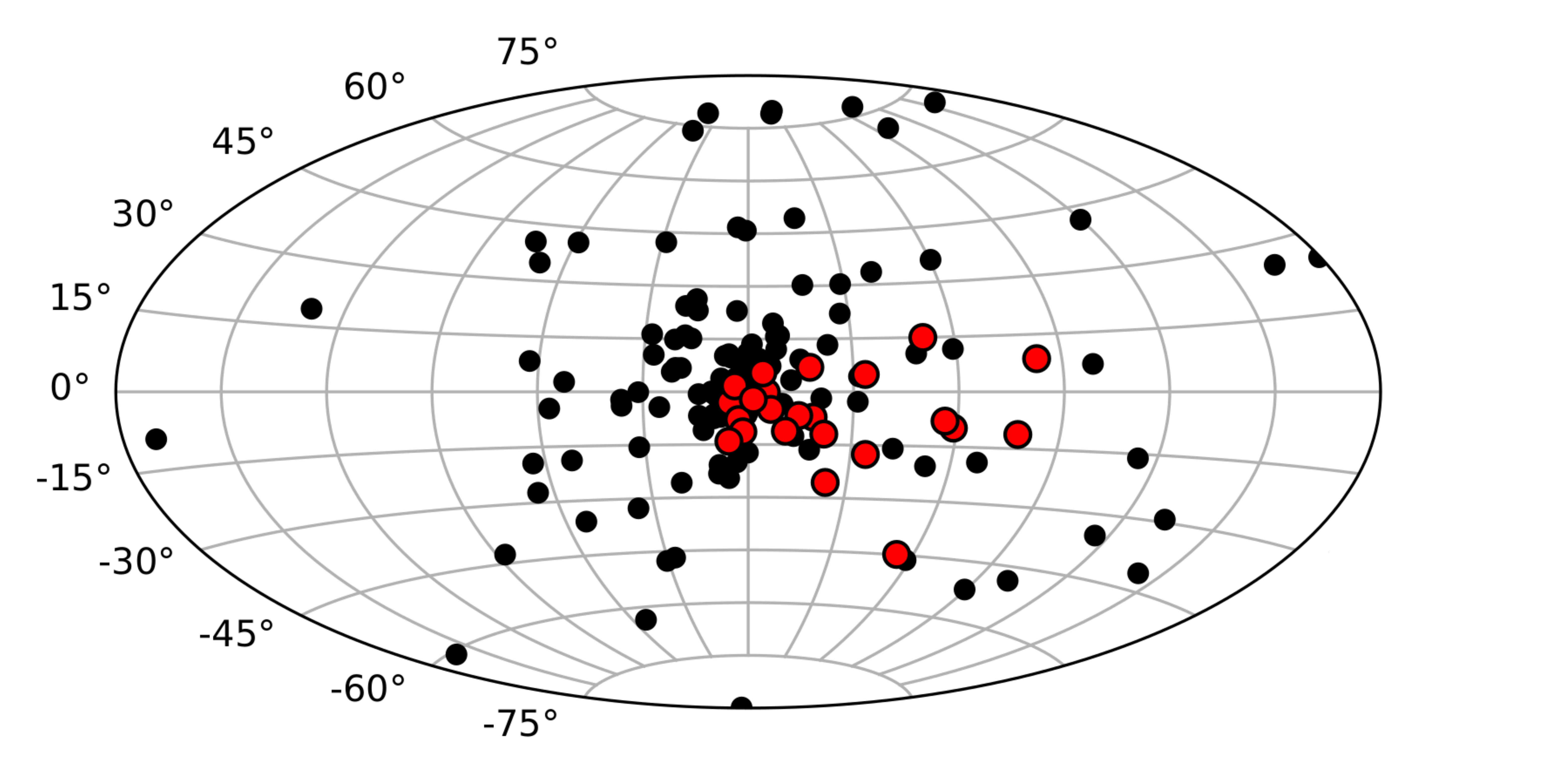}
	\caption{Galactic coordinate map in the Mollweide projection showing the distribution of clusters across the celestial sphere. The symbols are the same as in Figure~\ref{fig:sample}. The southern clusters are mostly visible in the Western Galactic hemisphere, and are concentrated towards the Galactic centre.}
	\label{fig:gal_pos}
\end{figure}

The more northerly clusters were observed with the Karl G. Jansky Very Large Array (VLA), with the results presented in \citet{Shishkovsky20}. The ATCA sample---presented in this paper---focused on southern Galactic globular clusters ($\delta < -25^\circ$). 
We retrieved most cluster properties (distances and radii) from the 2010 version of the Harris catalogue of Milky Way globular clusters\footnote{https://www.physics.mcmaster.ca/~harris/mwgc.dat} \citep{1996AJ....112.1487H}.  The properties of Liller\,1, Terzan\,1 and NGC\,6522 as listed in the Harris catalogue were inconsistent with their 2MASS images, so we updated their positions or sizes. For Liller\,1, we adopted the values of \citet{2015ApJ...806..152S}. Based on the 2MASS data, for Terzan\,1 we re-calculated the position, core and half-light radii, and of NGC\,6522 just the position. We use the cluster masses reported by \citet{2018MNRAS.478.1520B}. One cluster (NGC\,6652) in our sample was not covered by this study, so we used a mass-to-light ratio $\Upsilon = 2\,M_\odot L_\odot^{-1}$ to estimate the cluster mass based on its absolute visual magnitude ($M_{\rm V}$; \citealp{2005ApJS..161..304M}). In Figure~\ref{fig:sample} we show the masses and distances to our globular cluster sample, relative to the Galactic clusters in the \citet{1996AJ....112.1487H} catalogue. In Figure~\ref{fig:gal_pos}, we show the target distribution in Galactic coordinates---most of our sample is contained within $10^\circ$ from the Galactic centre. The cluster equatorial coordinates can be found in \citet{2018ApJ...862...16T}.

In Table~\ref{tab:gc_prop}, we list some basic properties for our sample which may correlate with the number of radio sources they contain. The stellar encounter rates have been estimated by \citet{2013ApJ...766..136B} using the integrated velocity dispersion and surface brightness profiles for most clusters. Where unavailable, we used their simpler estimates for NGC\,4833 (based on the core radius, surface brightness and velocity dispersion), and Liller 1 and Djorg 2 (based on the core radius and surface brightness). The number of known pulsars within each cluster is taken from the ATNF pulsar catalogue\footnote{\url{http://www.atnf.csiro.au/people/pulsar/psrcat/}}, but we caution that not all clusters have been surveyed for pulsars to the same depth. 

\begin{figure}
	\includegraphics[width=\columnwidth]{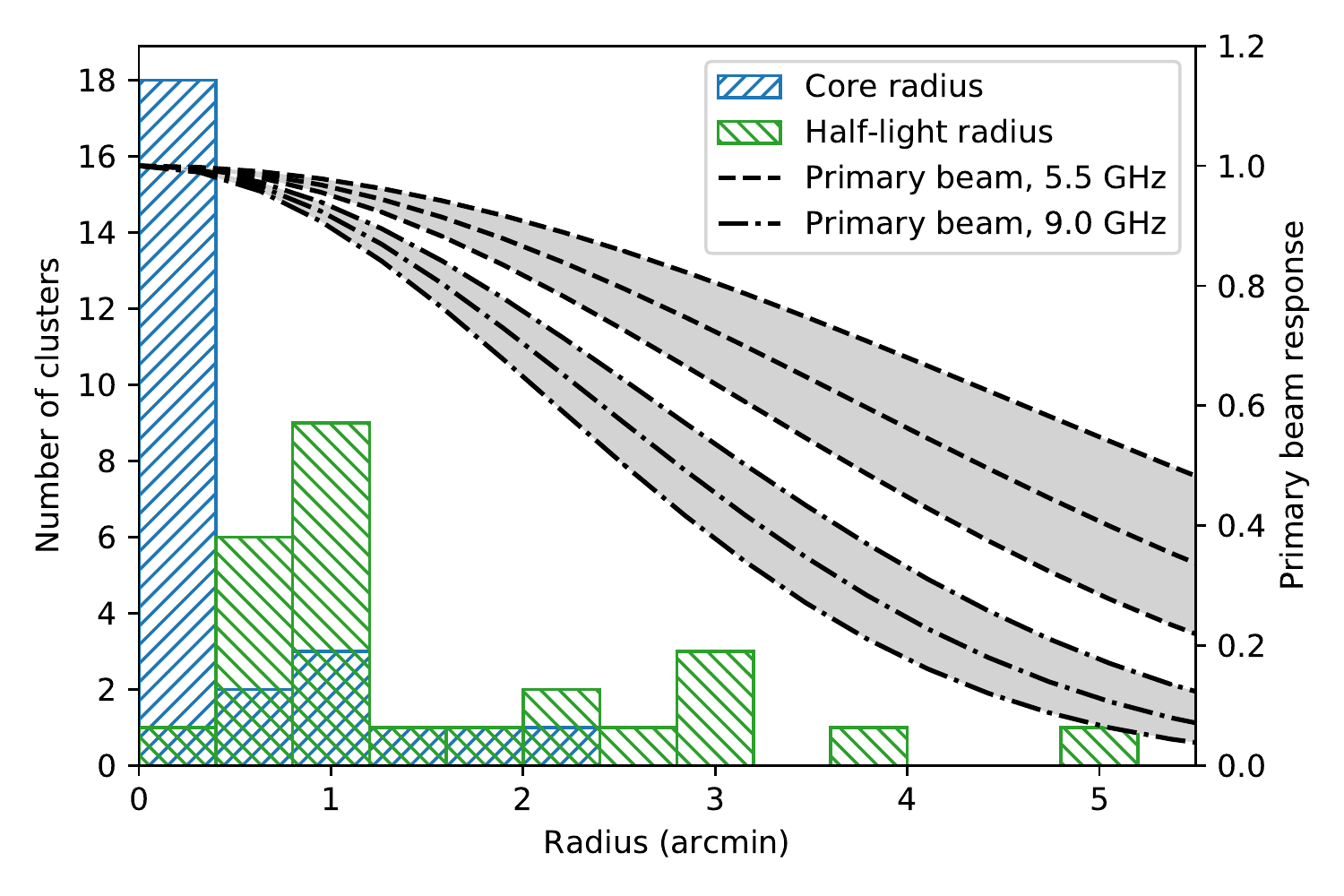}
	\caption{The distribution of core and half-light radii for the sample of 26 globular clusters, in relation to the primary beam responses at the observed frequencies of 5.5 GHz and 9.0 GHz. The primary beam response at a given radius and central frequency covers a band of values due to the large observing bandwidth (2\,GHz). Most clusters fit well within the primary beam, with the exception of $\omega$ Cen, the largest cluster plotted.}
	\label{fig:r_dist}
\end{figure}

\begin{table*}
	\caption{Properties of the globular clusters in the sample: the heliocentric distance ($D$), mass ($M$, from \citealt{2018MNRAS.478.1520B}), half-light radius ($R_{\rm h}$), core radius ($R_{\rm c}$), flag indicating whether the cluster is core-collapsed (cc), number of known pulsars (PSRs), and stellar encounter rate ($\Gamma$). Except where specified in Section~\ref{sec:selection}, distances and radii are taken from the 2010 version of the Harris catalogue \citep{1996AJ....112.1487H}, masses are taken from \citet{2018MNRAS.478.1520B}, encounter rates from \citet{2013ApJ...766..136B}, and pulsar numbers from the ATNF pulsar catalogue.}
	\centering
	\label{tab:gc_prop}
	\begin{tabular}{lSSSSSSr}
	\hline
	{Cluster} & {\shortstack{$D$\\(kpc)}} &  {\shortstack{$M$\\($10^5\,M_\odot)$}} & {\shortstack{$R_{\rm h}$\\(arcsec)}} & {\shortstack{$R_{\rm c}$\\(arcsec)}} & {\shortstack{cc}} & {\shortstack{Known\\PSRs}} & {$\Gamma$} \\
		\hline \hline
  {47 Tuc} &   4.6 &   7.8 &  190.2 &   21.6 &             &    27 &  1000  \\
 {Djorg 2} &   7.0 &   5.0 &   63.0 &   19.8  &             &     0 &    46  \\
 {Liller 1} &   8.1 &   6.7 &   30.5 &    5.4 &             &     0 &   391 \\
 {M54} &  23.9 &  14.1 &   49.2 &    5.4 &             &     0 &  2500  \\
 {NGC\,2808} &   9.4 &   7.4 &   48.0 &   15.0 &             &     0 &   923 \\
 {NGC\,3201} &   4.9 &   1.5 &  186.0 &   78.0 &             &     0 &   7 \\
 {NGC\,4372} &   6.3 &   2.5 &  234.6 &  105.0 &             &     0 &     0.2  \\
 {NGC\,4833} &   6.7 &   2.5 &  144.6 &   60.0 &             &     0 &    23  \\
 {NGC\,5927} &   7.9 &   3.5 &   66.0 &   25.2 &             &     0 &    68 \\
 {NGC\,6139} &  10.4 &   3.6 &   51.0 &    9.0 &             &     0 &   307  \\
 {NGC\,6304} &   5.9 &   2.8 &   85.2 &   12.6 &             &     0 &   123 \\
 {NGC\,6352} &   5.6 &   0.9 &  123.0 &   49.8 &             &     0 &     6.7  \\
 {NGC\,6362} &   7.6 &   1.5 &  123.0 &   67.8 &             &     0 &     4.6  \\
 {NGC\,6388} &  10.9 &  10.6 &   31.2 &    7.2 &             &     0 &   899  \\
 {NGC\,6397} &   2.3 &   0.9 &  174.0 &    3.0 &{\checkmark} &     2 &    84  \\
 {NGC\,6441} &  13.4 &  12.3 &   34.2 &    7.8 &             &     7 &  2300  \\
 {NGC\,6522} &   7.7 &   3.9 &   60.0 &    3.0 &{\checkmark} &     4 &   360  \\
 {NGC\,6541} &   7.5 &   2.8 &   63.6 &   10.8 &{\checkmark} &     0 &   390  \\
 {NGC\,6553} &   6.0 &   2.4 &   61.8 &   31.8 &             &     0 &    69 \\
 {NGC\,6624} &   8.4 &   0.7 &   49.2 &    3.6 &{\checkmark} &     10 &  1100  \\
 {NGC\,6652} &  10.0 &   0.8 &   28.8 &    6.0 &             &     2 &   700  \\
 {NGC\,6752} &   4.0 &   2.4 &  114.6 &   10.2 &{\checkmark} &     9 &   400 \\
 {$\omega$ Cen} &   4.9 &  35.5 &  300.0 &  142.2 &             &     5 &    90 \\
 {Terzan 1} &   5.2 &   2.2 &   23.8 &    2.0 &{\checkmark} &     7 &     0.3  \\
 {Terzan 5} &   5.9 &   5.7 &   43.2 &    9.6 &             &    39 &  6800 \\
 {Terzan 6} &   6.8 &   1.2 &   26.4 &    3.0 &{\checkmark} &     0 &  2500  \\
		\hline
	\end{tabular}
\end{table*}

\begin{table}
	\caption{Table of ATCA radio observations.}
	\fontsize{8}{9.5}\selectfont
	\centering
	\label{tab:rad_obs}
	\begin{tabular}{llll}
		\hline
		Source & Date & \shortstack{Time On Source (h)} & Phase calibrator \\
		\hline
		\hline
		47 Tuc			& 2010 Jan 24  $^{\rm a}$& 7.8 &	2353-686 \\
		        		& 2010 Jan 25  $^{\rm a}$& 10.5& 	\\
							& 2013 Nov 12  	& 8.7 & 	\\
							& 2015 Feb 02 	& 8.7 & 	\\
		\hline
		Djorg\,2			& 2015 Apr 26	& 8.1 & 1748-253\\
							& 2015 Apr 27	& 8.0 &	\\					
		\hline
		Liller\,1			& 2015 Jan 06	& 8.7 & 1714-336\\			
		\hline
		M54			& 2014 Dec 05	& 6.3 & 1921-293\\
		 				& 2014 Dec 06	& 5.0 & 	\\
		\hline
		NGC\,2808			& 2014 Dec 26	& 7.9 & J0845-6527\\
							& 2014 Dec 27	& 7.6 & \\
							& 2014 Dec 28	& 7.8 & \\
		\hline					
		NGC\,3201 			& 2014 Feb 05	& 9.3 & 1012-44	\\
							& 2014 Feb 06	& 8.8 & 	\\
		\hline
		NGC\,4372 			& 2015 Jan 04	& 8.4 & 1251-713\\
							& 2015 Jan 05	& 8.9 &	\\
		\hline
		NGC\,4833			& 2015 Jan 07	& 8.5 &	1251-713\\
							& 2015 Jan 08	& 8.4 & \\
		\hline
		NGC\,5927			& 2015 Jan 10	& 8.8 & 1511-47	\\
							& 2015 Jan 11	& 9.3 & 	\\		
		\hline
		NGC\,6139			& 2014 Dec 30	& 8.7 & 1606-39\\
							& 2014 Dec 31	& 4.9 & \\
							& 2015 Jan 01	& 3.0 & \\
		\hline
		NGC\,6304			& 2015 Jan 02	& 8.1 & 1710-269\\
							& 2015 Jan 03	& 8.3 & 	\\
		\hline	
		NGC\,6352 			& 2013 Nov 07	& 10.4 & 1740-517 \\
							& 2013 Nov 08	& 10.0 & 	\\
		\hline
		NGC\,6362			& 2014 Dec 02	& 9.9 & 1718-649 \\
							& 2014 Dec 05	& 5.1 & \\
							& 2015 Apr 23	& 10.1 & \\
		\hline
		NGC\,6388			& 2014 Dec 12	& 8.8 & 1714-397\\
							& 2014 Dec 13	& 9.2 & \\
							& 2014 Dec 14	& 9.1 & \\
		\hline
		NGC\,6397 			& 2013 Nov 09	& 8.8 & 1740-517 \\
							& 2013 Nov 10	& 6.9 & \\
		\hline
		NGC\,6441			& 2015 Apr 14	& 6.2 & 1729-37	\\
							& 2015 Apr 15	& 8.2 &	\\
							& 2015 May 02	& 8.1 &	\\
		\hline
		NGC\,6522			& 2015 Jan 16	& 8.5 & 1817-254\\
							& 2015 Jan 17	& 8.2 & 	\\
		\hline
		NGC\,6541			& 2015 Jan 14	& 8.1 & 1759-39	\\
							& 2015 Jan 15	& 8.3 & 	\\
		\hline		
		NGC\,6553			& 2015 Apr 30	& 7.9 & 1817-254 \\
							& 2015 May 01	& 8.4 & 	\\
		\hline
		NGC\,6624			& 2015 Apr 24	& 7.8 & 1817-254\\
							& 2015 Apr 25	& 8.5 & 	\\
		\hline
		NGC\,6652			& 2015 Apr 19 	& 8.5 & 1849-36\\
							& 2015 June 21	& 9.7 &	\\
							& 2015 June 22	& 7.9 &	\\
		\hline
		NGC\,6752			& 2014 Feb 06	& 2.6 & 1925-610\\
							& 2014 Feb 07	& 8.7 & \\
							& 2014 Feb 07	& 9.1 & \\
		\hline
		$\omega$ Cen			& 2010 Jan 22 $^{\rm a}$ & 10.1 & 1320-446\\
		& 2010 Jan 23 $^{\rm a}$ & 7.6 & \\
							& 2014 Dec 04	& 2.8 & \\
							& 2014 Dec 06	& 1.9 & 	\\
							& 2015 Jan 09	& 7.0 & 	\\
		\hline
		Terzan\,1			& 2015 Apr 20	& 8.2 & 1741-312\\
							& 2015 Apr 29	& 8.1 & \\		
		\hline
		Terzan\,5			& 2015 Apr 16 & 8.0 & 1748-253 \\
							& 2015 June 23 & 8.0 & \\
		\hline
		Terzan\,6			& 2015 Apr 17	& 8.0 & 1741-312 \\
							& 2015 Apr 18	& 7.9 & \\					
		\hline
		\multicolumn{4}{l}{\bf Notes.} \\
		\multicolumn{4}{l}{$^{\rm a}$ Archival data used by \citet{2011ApJ...729L..25L}.}
	\end{tabular}
\end{table}

\subsection{Setup}
The ATCA radio observations were carried out from 2013--2015 when the array was in any extended configuration (6\,km maximum baseline), primarily under project code C2877. We observed simultaneously in two bands, centred at 5.5 and 9.0\,GHz, each with a 2\,GHz bandwidth subdivided in $2048 \times 1$-MHz channels. We observed the source PKS\,1934--638 as a primary and bandpass calibrator for an integration time of $\approx$10\,minutes in each observing run. Secondary calibrators were observed for 1.5 or 2\,minute intervals each cycle depending on their flux density, and alternated with target observations of 5\,--\,20\,minute duration, with the cycle time depending on the atmospheric stability. Because ATCA is a six-element linear array of 22-m dishes, long exposure times are necessary both for good $uv$ coverage and for sensitivity. The on-source integration time per cluster varied between 8.7 and 35.7\,hr (median 16.8 hr), with a total of 499\,hr of on-source time, which also includes the archival data of \cite{2011ApJ...729L..25L} for 47 Tuc and $\omega$\,Cen (project code C2158). A summary of the radio observations is provided in Table~\ref{tab:rad_obs}.

In the part of the beam relevant for our study, the primary beam of ATCA is well described by a Gaussian profile with $FWHM \approx 9.2\arcmin$ and $5.4\arcmin$ at 5.5 and 9.0 GHz, respectively \citep{Wieringa92}. Excepting $\omega$\,Cen, the half-light diameters of all our sample globular clusters fit within the FWHM of ATCA at 5.5 GHz (Figure~\ref{fig:r_dist}), and the core radii are within the FWHM at both frequencies. Given the likelihood of mass segregation for compact objects, we expect that most radio continuum sources associated with the globular clusters should be within our ATCA fields of view.

\subsection{Data reduction}
Data reduction and calibration were performed using the \textsc{Miriad} v1.5 \citep{1995ASPC...77..433S} software package using standard procedures. We calibrated the 5.5 and 9.0 GHz data separately. A stack of visibilities between the two frequencies was also performed to increase our sensitivity to flat-spectrum point sources. As such, the stacked data has a lower root-mean-square (rms) noise than either of the individual bands, at the cost of being less sensitive to steep-spectrum sources such as pulsars. We used pixel sizes of 0.33\arcsec{}, 0.24\arcsec{} and 0.24\arcsec{} for the 5.5, 9.0 and 7.25 GHz images respectively. We imaged and deconvolved the Stokes I data in \textsc{CASA} v5.1.0 \citep{2007ASPC..376..127M} using the multi-frequency, multi-scale \textsc{clean} \citep{1980A&A....89..377C, 1994A&AS..108..585S,2011A&A...532A..71R} algorithm with 2 Taylor coefficients, which takes into account source spectral indices; and Briggs weighting of robustness 1 \citep{briggs95}, which gives a good compromise between sensitivity and spatial resolution. We imaged the 5.5 and 9.0 GHz data and their stack (7.25 GHz) separately. To improve dynamic range, self-calibration can be employed to better solve for time-variable atmospheric phases in the direction of the target. Due to signal--to--noise considerations, self-calibration was only possible for fields containing a total flux density greater than $\sim$3\,mJy . We performed this step for five fields: Djorg\,2, M54, NGC\,6441, NGC\,6624, and NGC\,6752. We cropped all images within the 20\% primary beam response, and corrected for frequency-dependent primary beam attenuation using the task {\tt widebandpbcor}. For cropping the stacked image, we used the 20\% response of the 9\,GHz primary beam. This resulted in image radii of 6.6\arcmin{}, 4.2\arcmin{} and 4.2\arcmin{} at 5.5, 9.0 and 7.25\,GHz respectively. The achieved resolutions and sensitivities are listed in Table~\ref{tab:img-props}.

\begin{table*}
\centering
\fontsize{9}{10.5}\selectfont
\caption{The time on source ($t_{\rm obs}$) for each cluster, together with the FWHM of the synthesised beam -- major axis ($B_{\rm maj}$), minor axis ($B_{\rm min}$) and position angle ($B_{\rm pa}$, measured from north toward east), and the central rms, for each of the three frequencies -- 5.5\,GHz, 9\,GHz, and their stack (7.25\,GHz). Approximately half of the data on 47 Tuc and $\omega$ Cen were taken by \citet{2011ApJ...729L..25L}, prior to the 4-cm receiver upgrade, and are therefore less sensitive.}
\label{tab:img-props}
\begin{tabular}{cc|cccc|cccc|cccc|}
\cline{3-14}
 & & \multicolumn{12}{c|}{Frequency} \\ \cline{3-14} 
 & & \multicolumn{4}{c|}{5.5 GHz} & \multicolumn{4}{c|}{9 GHz} & \multicolumn{4}{c|}{7.25 GHz} \\ \hline
\multirow{2}{*}{Source}  
 & $t_{\rm obs}$ & $B_{\textrm{maj}}$ & $B_{\textrm{min}}$ & $B_{\textrm{pa}}$ & rms & $B_{\textrm{maj}}$ & $B_{\textrm{min}}$ & $B_{\textrm{pa}}$ & rms & $B_{\textrm{maj}}$ & $B_{\textrm{min}}$ & $B_{\textrm{pa}}$ & rms\\
 & (hr) & (\arcsec) & (\arcsec) & (deg) & (\si{\micro}Jy\,bm$^{-1}$) & (\arcsec) & (\arcsec) & (deg) & (\si{\micro}Jy\,bm$^{-1}$) & (\arcsec) & (\arcsec) & (deg) & (\si{\micro}Jy\,bm$^{-1}$)  \\ \hline \hline
47 Tuc  &  35.7 & 2.2 &  1.7 &  32 &  3.8 &  1.5 &  1.1 &  39 &  4.7 &   1.9 &  1.4 &  32 &  3.2 \\
Djorg 2  &  16.1 & 4.8 &  1.5 &   0 &  4.2 &  3.1 &  1.0 &  -1 &  4.2 &   3.8 &  1.2 &   0 &  3.2 \\
Liller 1 &  8.7 &  3.6 &  1.6 &   1 &  6.2 &  2.5 &  1.1 &   0 &  6.1 &   3.0 &  1.3 &   0 &  4.4 \\
M54 &  11.3 &  4.4 &  1.6 & -12 &  5.3 &  3.1 &  1.4 & -20 &  9.7 &   3.9 &  1.5 & -14 &  5.7 \\
NGC\,2808 &  23.3 &  2.6 &  1.5 &   9 &  3.7 &  1.7 &  1.0 &   9 &  4.3 &   2.0 &  1.3 &  13 &  2.9 \\
NGC\,3201 &  18.1 &  2.9 &  1.5 &   4 &  3.9 &  2.2 &  1.1 &   4 &  4.1 &   2.2 &  1.1 &   4 &  3.5 \\
NGC\,4372 &  17.3 &  2.2 &  1.6 &   0 &  4.1 &  1.5 &  1.1 &   0 &  4.6 &   1.8 &  1.3 &   0 &  3.0 \\
NGC\,4833 &  16.9 &  2.3 &  1.6 &  -3 &  4.1 &  1.6 &  1.1 &  -3 &  4.5 &   1.9 &  1.3 &  -4 &  3.0 \\
NGC\,5927 &  18.1 &  2.6 &  1.7 &   0 &  4.2 &  1.7 &  1.2 &  -1 &  5.3 &   2.1 &  1.4 &  -1 &  3.3 \\
NGC\,6139 &  16.6 &  2.9 &  1.8 &   9 &  4.2 &  2.0 &  1.3 &   6 &  4.9 &   2.3 &  1.5 &   6 &  3.3 \\
NGC\,6304 &  16.4 &  4.1 &  1.6 &   0 &  3.9 &  2.8 &  1.1 &   0 &  4.2 &   3.3 &  1.3 &   0 &  2.9 \\
NGC\,6352 &  20.4 &  2.6 &  1.7 &   0 &  3.6 &  1.8 &  1.2 &   7 &  4.3 &   2.1 &  1.4 &   3 &  2.8 \\
NGC\,6362 &  25.1 &  2.1 &  1.7 &  11 &  3.3 &  1.4 &  1.2 &   5 &  3.9 &   1.7 &  1.4 &   6 &  2.5 \\
NGC\,6388 &  27.1 &  3.2 &  1.8 &  -2 &  3.5 &  2.0 &  1.2 &  -7 &  3.4 &   2.5 &  1.4 &  -4 &  2.6 \\
NGC\,6397 &  15.7 &  2.9 &  1.5 &   0 &  4.6 &  1.9 &  1.0 &  -3 &  4.8 &   2.2 &  1.2 &  -2 &  3.3 \\
NGC\,6441 &  22.5 &  3.8 &  1.5 &   2 &  3.5 &  2.4 &  1.0 &   4 &  3.4 &   2.8 &  1.2 &   4 &  2.6 \\
NGC\,6522 &  16.7 &  4.0 &  1.6 &  -1 &  4.1 &  2.7 &  1.1 &  -1 &  4.3 &   3.2 &  1.3 &  -1 &  3.0 \\
NGC\,6541 &  16.4 &  3.0 &  1.6 &  -2 &  5.4 &  2.1 &  1.1 &  -2 &  4.6 &   2.7 &  1.5 &  -2 &  3.7 \\
NGC\,6553 &  16.3 &  5.0 &  1.6 &   1 &  4.1 &  3.4 &  1.1 &   1 &  4.3 &   4.0 &  1.3 &   1 &  3.0 \\
NGC\,6624 &  16.3 &  4.2 &  1.6 &   0 &  4.0 &  2.8 &  1.1 &   0 &  4.2 &   3.3 &  1.3 &   0 &  3.0 \\
NGC\,6652 &  26.1 &  3.7 &  1.5 &  -1 &  3.3 &  2.7 &  1.1 &   0 &  3.5 &   3.1 &  1.3 &  -1 &  2.5 \\
NGC\,6752 &  20.4 &  2.8 &  1.5 & -15 &  3.8 &  1.8 &  1.0 & -17 &  4.3 &   2.3 &  1.3 & -17 &  2.9 \\
$\omega$ Cen&  29.4 &  3.0 &  1.6 & -16 &  3.7 &  2.0 &  1.1 & -18 &  4.5 &   2.4 &  1.3 & -18 &  2.9 \\
Terzan 1 &  16.3 &  4.2 &  1.5 &   0 &  5.6 &  2.9 &  1.1 &  -1 &  4.1 &   3.4 &  1.3 &   0 &  2.9 \\
Terzan 5 &  16.0 &  5.1 &  1.5 &   1 &  5.1 &  3.5 &  1.0 &  -1 &  6.5 &   4.2 &  1.3 &   0 &  4.9 \\
Terzan 6 &  15.9 &  4.2 &  1.6 &   3 &  4.7 &  2.8 &  1.1 &   4 &  5.1 &   3.4 &  1.3 &   4 &  3.6 \\
\hline
\end{tabular}
\end{table*}

\section{Methods}
\label{sec:method}

\subsection{Source finding}
\label{sec:src_find}

We use the source extraction software {\sc pybdsf} v1.8.13 \citep{2015ascl.soft02007M} to find radio sources in the 5.5, 9.0, and the (stacked) 7.25 GHz images. This tool estimates the the local background by measuring the mean and rms using a sliding box algorithm. We set the size of the sliding box to $150 \times 150$ pixels, measured in steps of 30 pixels. For bright sources of signal-to-noise ratio (SNR) $\geq 50$, around which imaging artefacts are more likely, we use a smaller box of $50\times50$ pixels in steps of 10 pixels. 

Sources are then found by identifying pixels with values larger than three times the rms above the mean background; these are the seeds of islands of emission. Island boundaries are expanded to include adjacent pixels with values larger than 3$\sigma$, and are then fit with one or more Gaussians fixed to the shape of the synthesised beam. We assume each contiguous island of emission is a single source, with a flux density equal to the modelled total flux density (for point sources, this is equal to the peak flux density of the Gaussian). We note that this approach may improperly carve up extended sources with patchy islands of bright emission into multiple sources, but in practice this seldom appears to occur in our catalogue. In any case, we expect true globular cluster sources to mostly be point sources, with perhaps rare exceptions such as known planetary nebulae.

Because the response of the primary beam decreases quickly towards its edge, we estimated the rms and background maps from the non-primary-beam-corrected images, and then corrected them for the primary beam response. Source finding was run on the primary-beam-corrected images. Since we cropped all images at the 20\% primary beam response, the uncertainty in the primary beam model only minimally affects the flux density measurements. Primary beam effects are even less important for sources within globular clusters, as these are typically found close to the phase centre. At these frequencies, systematic errors in the absolute flux density scale are expected to be at the level of $\sim 2$ per cent \citep{Reynolds94}. Hence, for most sources, we expect flux density errors to be dominated by signal-to-noise constraints. We therefore report the flux density errors as statistical errors from the fit. 

After source-finding, we cross-match the catalogues at the three different frequencies. We consider a match any two sources (at different frequencies) whose 3$\sigma$ positional uncertainty ellipses overlap. A source may not have a counterpart at two frequencies because: (a) it lies outside of the 20\% primary beam response at the higher frequency, (b) any signal is buried by the noise, or (c) the morphology of an extended source is significantly different between each image. In case (a), we do not report the flux density of any source missing from the 9\,GHz or 7.25\,GHz catalogue (the primary beams at these frequencies are smaller than at 5.5\,GHz). In case (b), we take the non-detection as an upper limit at $3\times {\rm rms}$. In case (c), we take the value of the upper limit as the value of the pixel at the position of the source, plus $3\times {\rm rms}$.

In total, six cross-matching steps are performed. For each source in the 5.5\,GHz catalogue, a match (or upper limit) is searched for in the 7.25\,GHz catalogue and in the 9\,GHz catalogue. After these two steps, each source in the 5.5\,GHz catalogue is fully characterised, but there are remaining sources in the 7.25\,GHz and 9\,GHz catalogues that still need to be described - those sources that have been detected at higher frequencies, but not at 5.5\,GHz. For each of these sources in the 7.25\,GHz catalogue we find matches (or upper limits) at 5.5\,GHz and 9\,GHz. Similarly, the same process is carried out for leftover sources at 9\,GHz. 

After cross-matching between the three catalogues, we only keep those sources with SNR $\geq 5$ in at least one band, to ensure the catalogue is reliable. We measure the spectral index ($S_\nu \propto \nu^\alpha$) as $\alpha = \log(S_{5.5}/S_{9}) / \log(5.5/9)$. If a radio source is only visible in one band, we can place a limit on the spectral index by fixing the value of the flux density in the other band to its upper limit. Errors on the spectral index are estimated by resampling the flux density measurements assuming Gaussian errors, and recalculating the spectral index 1000 times for each source. As the differences between the negative and positive errors of spectral indices do not typically exceed values of $\sim$0.1, we report symmetric spectral index errors as the geometric mean of the two errors.
We note that a small number of the spectral indices have unphysically extreme values: these are predominantly extended background sources that have different amounts of emission resolved out at 5.5 and 9 GHz. 

\subsection{Source injection}

\begin{figure}
	\centering
\includegraphics[height=0.75\textheight]{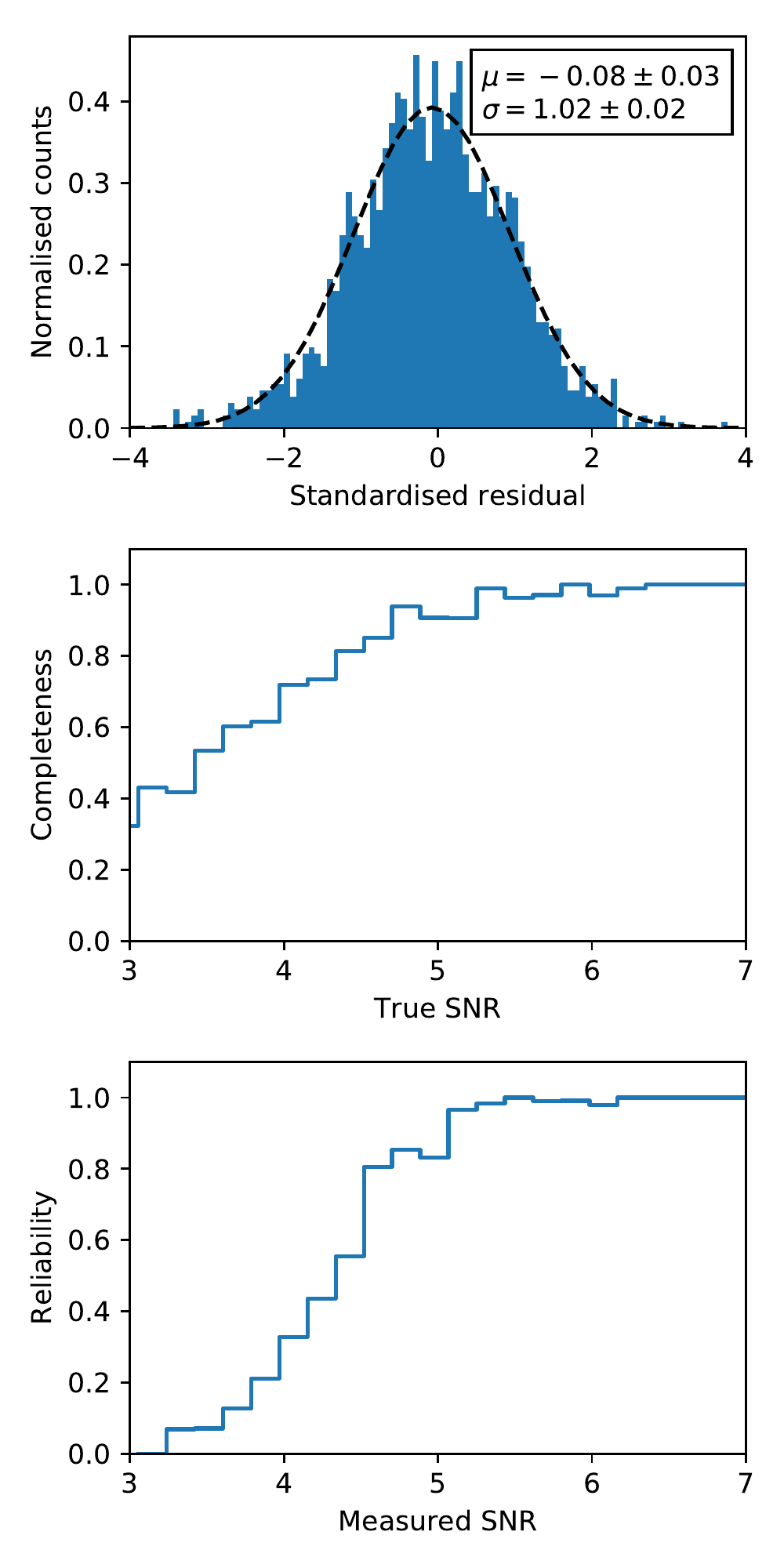}
	\caption{Source recovery statistics. {\it Top:} normalised histogram of standardised residuals for the matched sources, $(S_{\rm true} - S_{\rm obs}) / \sigma_{S_{\rm obs}}$, for sources with measured SNR $> 5$. The dashed black line shows the normal distribution of best fit. {\it Middle:} completeness as a function of input SNR. Sources are virtually complete above $5\sigma$. {\it Bottom:} reliability as a function of measured SNR. Above $5\sigma$, sources are reliably identified.}
	\label{fig:stats}
\end{figure}

We performed a test to assess whether the completeness, reliability, and flux recovery performance of {\sc Pybdsf} using the method described in Section~\ref{sec:src_find} are optimal. This consisted of simulating a simple radio observation by injecting 2000 sources into one of our measurement sets (of NGC\,2808), into a 500\arcsec$\times$500\arcsec\ field 1 deg away from the phase centre, to suppress any real sources. We injected these sources at positions at least 7\arcsec\ away from each other, with flux densities corresponding to a SNR in the range 3 to 8. The rms is 4.2\,\si{\micro}Jy\,beam$^{-1}$, and the beam size is 1\arcsec$\times$1.75\arcsec. After imaging the simulated field, we extracted the sources as described in the previous section, and cross-matched them with the true source catalogue. In Figure~\ref{fig:stats}, we show the statistics associated with source extraction. We find that across the recovered sources, the standardised residuals, $(S_{\rm true} - S_{\rm obs}) / \sigma _{S_{\rm obs}}$, have a mean $\mu = -0.31$ and standard deviation $\sigma = 1.01$. The measured flux densities are over-estimated (expected $\mu = 0$), as a form of Eddington bias \citep{1913MNRAS..73..359E}, which makes faint sources more likely to be detected if they coincide with noise peaks. For sources with true SNR $> 5 \sigma$, this bias is minimal ($\mu = -0.08$). Above this significance level, we find the completeness and reliability to be above the 90\% level. For these reasons, we choose to set the threshold for defining a detection at $5\sigma$.

\section{Results}
\label{sec:result}

We detected a total of 1285 sources across our 26 globular cluster fields. 352 of these have counterparts at both 5.5 and 9.0\,GHz and hence for which a spectral index
could be calculated. 36 sources are located within one core radius, and 189 sources within one half-light radius; these are the sources most likely to be truly associated with the globular clusters. We detail the characteristics of these sources in our source catalogue, which contains the following information:
\begin{enumerate}
	\item Source ID	(Cluster-ATCA\#\#, sorted by distance from cluster centre)
	\item Right ascension (RA)
	\item Declination (DEC)
	\item 1$\sigma$ error in RA ($\Delta$\,RA)
	\item 1$\sigma$ error in DEC ($\Delta$\,DEC)
	\item Flux density at 5.5\,GHz ($S_{5.5}$)
	\item 1$\sigma$ error in $S_{5.5}$ ($\Delta \,S_{5.5}$)
	\item Flux density at 7.25\,GHz ($S_{7.25}$)
	\item 1$\sigma$ error in $S_{7.25}$ ($\Delta \,S_{7.25}$)
	\item Flux density at 9\,GHz ($S_{9}$)
	\item 1$\sigma$ error in $S_{9}$ ($\Delta \,S_{9}$)
	\item Spectral index $\alpha$
	\item 1$\sigma$ error in $\alpha$
	\item Distance from the centre in units of core radii ($R_c$)
	\item Distance from the centre in units of half-light radii ($R_h$)
\end{enumerate}

\begin{figure*}
    \centering
    \includegraphics[width=\textwidth]{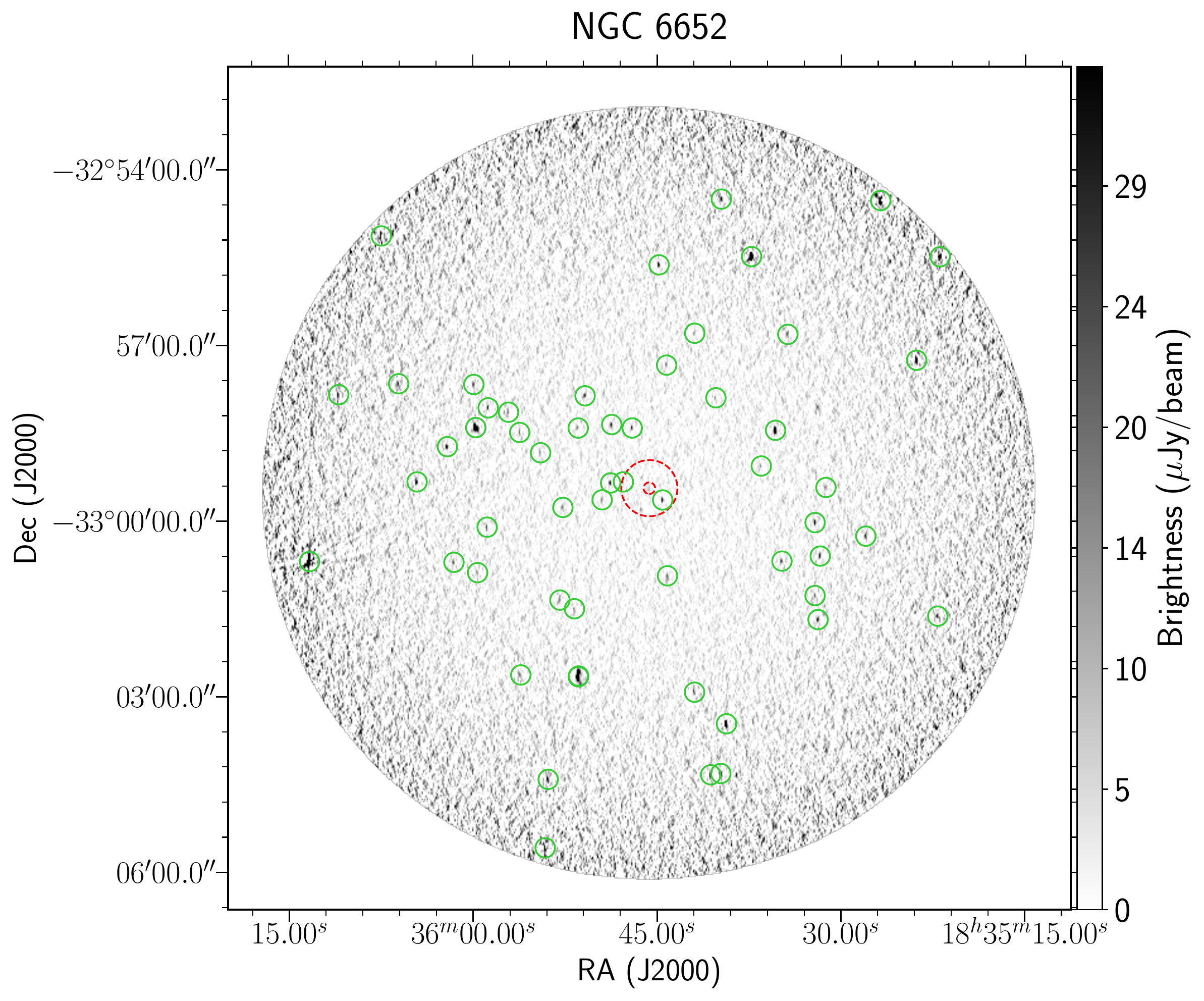}
    \caption{5.5 GHz image of the NGC\,6652 field. The core and half-light radius of the cluster are shown in red dotted circles, and the detected sources are highlighted with green circles.}
    \label{fig:n6652}
\end{figure*}

We show the first 20 lines from the source catalogue in Table~\ref{tab:source-cat}.
Figure~\ref{fig:n6652} gives an example image from the survey with the catalogued sources overlaid. The complete source catalogue and the final images can be downloaded from the CSIRO Data Access Portal \footnote{\url{https://doi.org/10.25919/nzcz-xv90}}.

\begin{table*}
\fontsize{7}{8.5}\selectfont
\caption{The first 20 sources in the source catalogue. The full table is available from \url{https://data.csiro.au/collection/csiro:54270}.}
\label{tab:source-cat}
\include{47Tuc_cat}

\flushleft {\bf Notes.} \\
$^{\rm a}$ The quoted errors in astrometry are statistical only. An additional systematic error of order 1/10th of the beam size might be expected (see discussion in  \citealt{Shishkovsky20}).
\end{table*}

\subsection{Source counts}

Radio source counts ($\log N/\log S$ as a function of $\log S$) have been used since the early days of radio astronomy as a tool for exploring the evolution of the Universe \citep{1966MNRAS.133..421L}. High flux density sources are mostly active galactic nuclei (AGN), but star-forming galaxies dominate at the faint end \citep{2008MNRAS.386.1695S, 2012ApJ...758...23C}. The different populations of astrophysical sources can be seen as a flattening of the source counts below 1\,mJy \citep{1984A&AS...58....1W}. Sample variance (due to the anisotropic distribution of sources across the sky) can induce deviations in the radio source counts measured in a given survey, depending on flux density and survey area \citep{2013MNRAS.432.2625H}.

In this paper, we use radio source counts as a tool to identify the putative excess of radio sources towards globular clusters. Because differential, rather than cumulative, source counts are less liable to biases \citep{1970ApJ...162..405C}, we employ this metric. The number of sources per unit flux density and unit sky area, known as differential source counts, $n(S_j)$, is given by \citep{2013MNRAS.428.3509E}:
\begin{equation}
n(S_j) = \frac{dN}{dS d\Omega} = \frac{\sum_{i = 1}^{i = m} \frac{1}{\Omega_i}}{\Delta S_j},
\end{equation}
where $j$ is the flux bin between $S_0$ and $S_1$ of mean $S_j$ and width $\Delta S_j = S_1 - S_0$, which contains $m$ sources. $\Omega_i$ is the sky coverage that is sensitive enough to detect sources of flux density $S_i$ with at least $5\sigma$ confidence. We take $S_j$ to be the geometric mean $\sqrt{S_0 S_1}$.

To estimate the errors on the source counts, we first investigate how dominant the sources of error are: Poisson errors, errors in the sensitivity map, or flux density errors. The fractional error of the rms is $\sigma_{\rm rms} / {\rm rms} = (1/2N_{\rm beams})^{1/2}$ \citep{2012ApJ...758...23C}, where $N_{\rm beams}$ is the number of resolution elements (synthesised beams) within an area. For a $100 \times 100$ pixel box, $N_{\rm beams} \approx 60$, so the error on the rms is at the 10\% level. For flux density bin widths (we use 0.1\,dex, i.e. a width of 5\,\si{\micro}Jy for our lowest flux density bin centred at 20\,\si{\micro}Jy) that are almost all much larger than typical flux density errors for individual sources ($\approx 4$\,\si{\micro}Jy), the errors in the source counts associated with measurement errors are negligible. Therefore, for low number counts ($< 100$ counts bin$^{-1}$), the errors in the measured source counts are dominated by Poisson errors.

\begin{figure}
	\begin{subfigure}[t]{\columnwidth}
		\includegraphics[width=\textwidth]{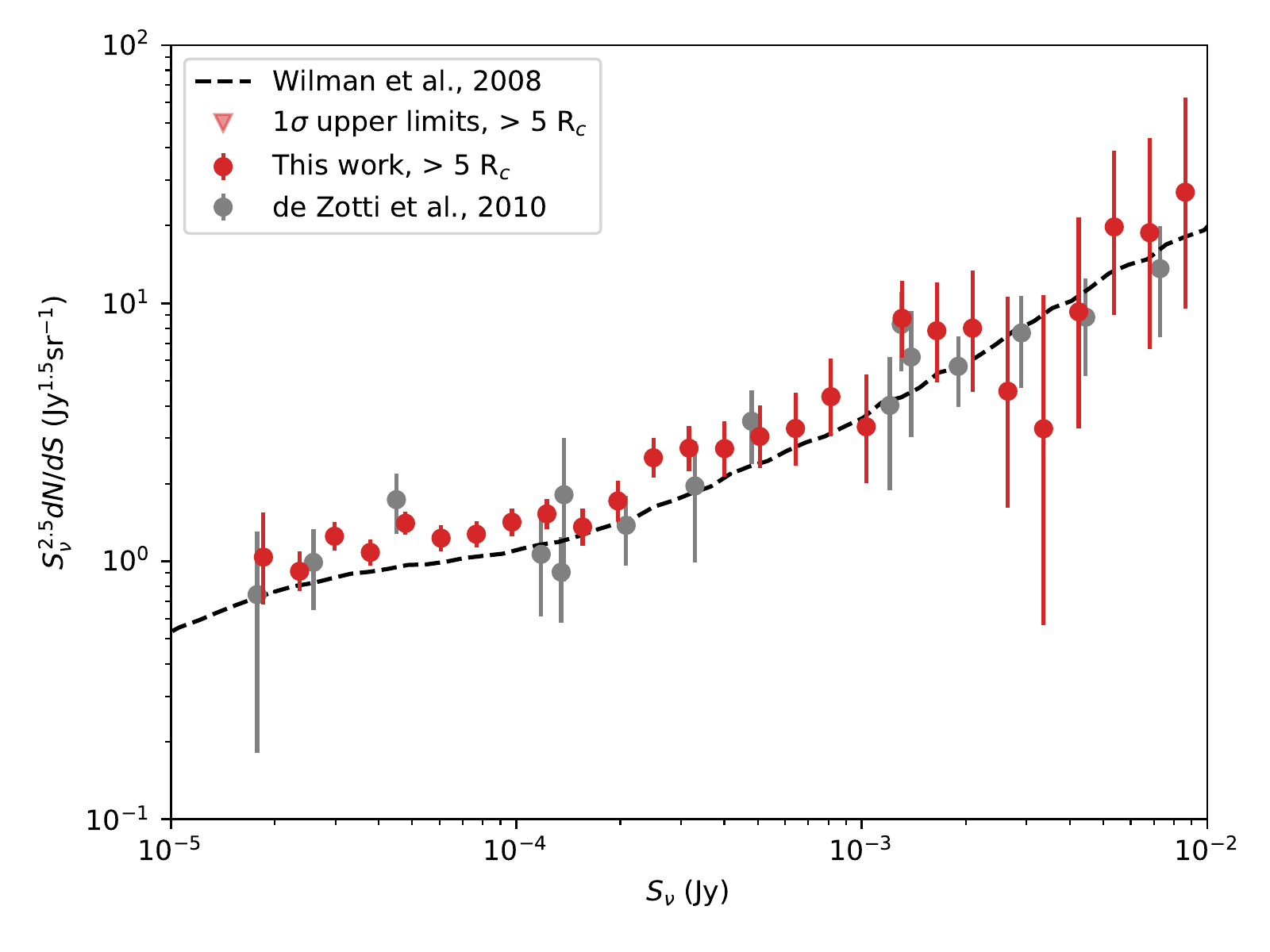}
		\caption{Source counts at 5.5\,GHz}
		\label{fig:sc_out_5p5}
	\end{subfigure}
	\begin{subfigure}[t]{\columnwidth}
		\includegraphics[width=\textwidth]{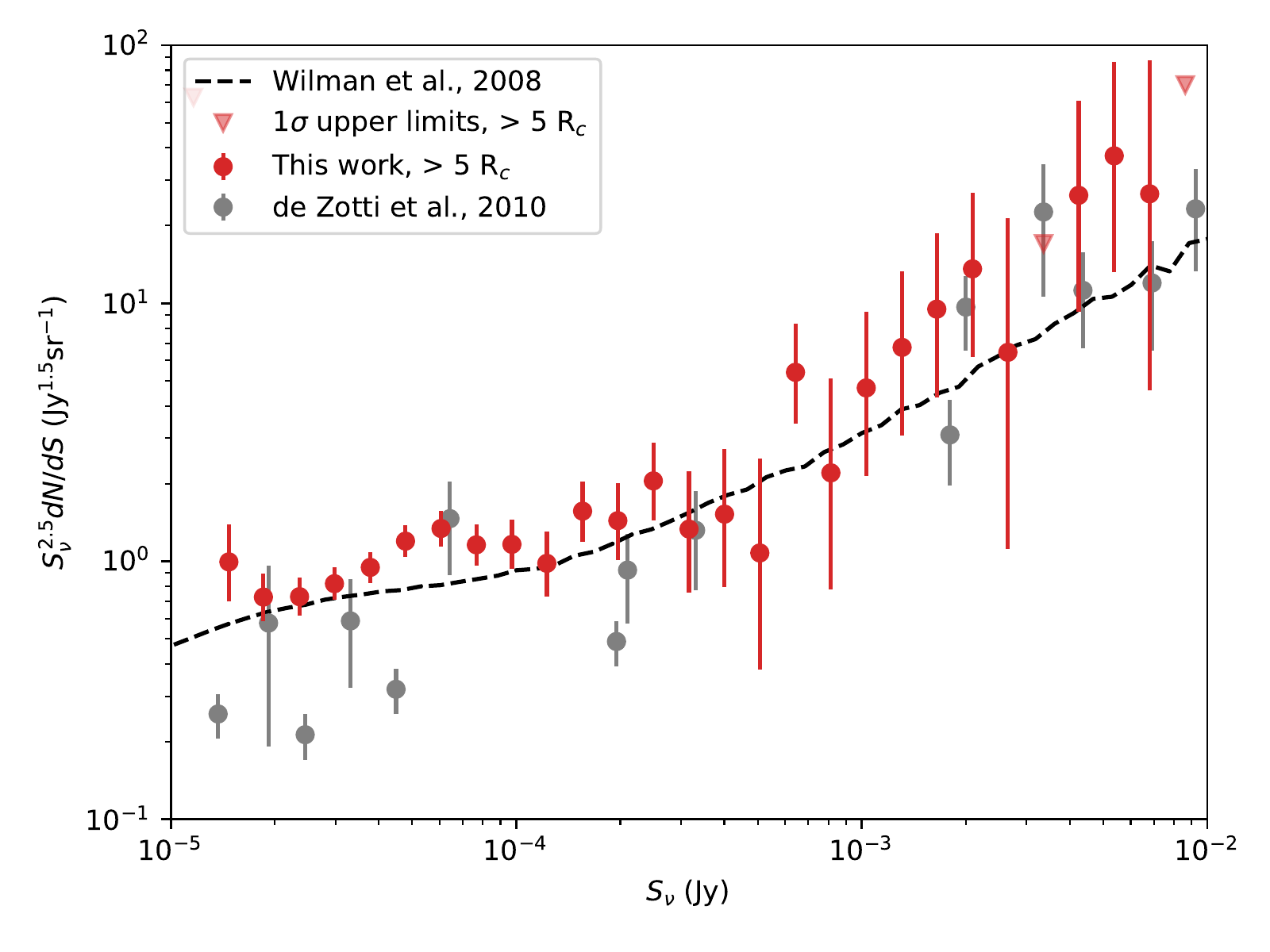}
		\caption{Source counts at 7.25\,GHz}
		\label{fig:sc_out_7p25}
	\end{subfigure}
	\centering
	\begin{subfigure}[t]{\columnwidth}
		\includegraphics[width=\textwidth]{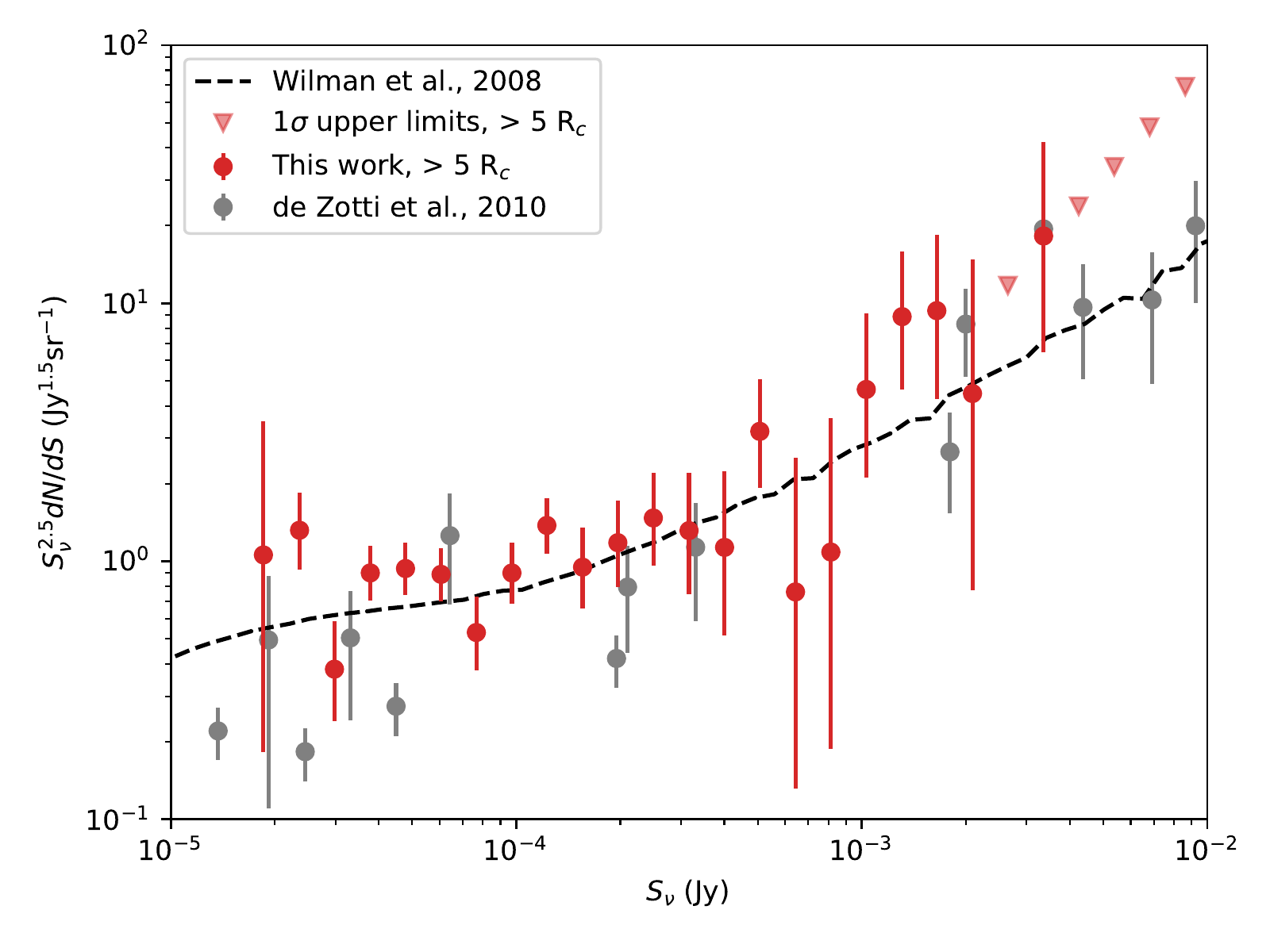}
		\caption{Source counts at 9.0\,GHz}
		\label{fig:sc_out_9p0}
	\end{subfigure}
	\caption{Differential source counts outside clusters ($>5 R_c$) at three different frequencies, compared to the collated measurements of \citet{2010A&ARv..18....1D}, and the theoretical estimates of \citet{2008MNRAS.388.1335W}. The error bars represent 1$\sigma$ uncertainties, and the triangles represent 1$\sigma$ upper limits. The source counts derived from our work are consistent with previous estimates within a factor of 2. Disparities are due to observational and processing biases rather than cosmic variance.}
	\label{fig:sc_out}
\end{figure}

In Figure~\ref{fig:sc_out}, we show the source counts measured from all of our sampled fields outside of clusters, at three different frequencies, and compare them with the previous estimates of the radio source counts of \citet{2010A&ARv..18....1D} and \citet{2008MNRAS.388.1335W}. \citet{2010A&ARv..18....1D} provide a collation of source counts measured from different surveys at various frequencies, and with different sensitivities. \citet{2008MNRAS.388.1335W} performed a semi-empirical simulation of the radio sky covering $20\times20$\,deg$^2$ down to 10\,nJy at four different frequencies. To convert the observed source counts of \citet{2010A&ARv..18....1D} to our observing frequencies, we rescaled their 4.86\,GHz source counts to 5.5\,GHz, and their 8.4\,GHz source counts to 7.25 and 9.0\,GHz, assuming a constant spectral index $\alpha = -0.7$ \citep{1984ApJ...287..461C}. We calculated the theoretical source counts from \citet{2008MNRAS.388.1335W} first by measuring the spectral index for each source in their catalogue using the total flux densities at 4.86\,GHz and 18\,GHz, and then interpolating to 5.5, 7.25 and 9.0\,GHz. To mitigate the contamination with globular cluster sources for our sample, we have excluded the regions within $5 R_c$ of cluster centres. 

\subsection{Determining an excess of sources toward globular clusters}

Many or most of the radio sources in our catalogue are likely associated with 
the background rather than the targeted clusters. Detailed studies have already identified several such extragalactic background sources, including N3201-ATCA1 and N3201-ATCA5 \citep[within the core radius of NGC3201;][]{Paduano22}, and likely all other sources within the half-light radius of that cluster; N6397-ATCA9, N6397-ATCA13, N6397-ATCA14 and N6397-ATCA18 \citep[X-ray sources U108, W129, W127, and W135 towards NGC\,6397, respectively;][]{Zhao20}; N6752-ATCA3, N6752-ATCA5 and N6752-ATCA7 \citep[X-ray sources CX42, CX17, and CX45 towards NGC\,6752, respectively;][]{Cohn2021}, as well as all sources outside the half-light radius of Terzan 5 \citep{Urquhart20}.

Nonetheless, several of our sources are known to be associated with the target clusters. Here we present a preliminary analysis of how many catalogue sources are indeed cluster members.

\subsubsection{Known Sources}

The first and most definitive check is whether a source is already a known member of a globular cluster.

Six radio sources in our catalogue are associated with previously known pulsars with published positions: in 47 Tuc (pulsar D, \citealt{1995MNRAS.274..547R}; associated with 47Tuc-ATCA2), in NGC\,6397 (pulsar A, \citealt{2001ApJ...548L.171D}; associated with N6397-ATCA4), in NGC\,6522 (pulsar A, \citealt{2020ApJ...905L...8Z}; associated with N6522-ATCA1), and in Terzan\,5 (pulsars P, A and C, \citealt{1990Natur.347..650L,2005Sci...307..892R};  associated with Ter5-ATCA2, Ter5-ATCA4, and Ter5-ATCA5, respectively; see also \citealt{Urquhart20}). These sources generally have strongly negative spectral indices. It is nearly certain that other pulsars are present in our catalogue: for example, \citet{Zhao20} showed that a steep-spectrum radio source near the centre of NGC\,6397 (here called N6397-ATCA1) is associated with the X-ray/optical source U18 and is very likely to be a ``spider" binary pulsar where the pulsations are frequently eclipsed by material lost from the redback companion.

Apart from pulsars, we have also detected some well-known X-ray binaries---in Terzan\,5, EXO 1745--248, which was in outburst at the time of the observations \citep{2016MNRAS.460..345T}, here catalogued as Ter5-ATCA3; in NGC\,6624, 4U\,1820--30, here catalogued as N6624-ATCA1 (see \citealt{2021MNRAS.508L...6R} and \citealt{Panurach21}); and in Terzan\,6, GRS 1747--312, here catalogued as Ter6-ATCA1 \citep[see][]{Panurach21}. We have previously published two X-ray binaries discovered as part of this ATCA survey: in 47\,Tuc, the candidate black hole X9 \citep{2015MNRAS.453.3918M}, here catalogued as 47Tuc-ATCA1; and in NGC\,6652, the candidate transitional millisecond pulsar NGC6652B \citep{2001ApJ...562..363H,Paduano21}, here catalogued as N6652-ATCA3.

We emphasize that while a few detailed studies of individual clusters have been published \citep[see, e.g.,][]{Zhao20,Urquhart20,Cohn2021,Paduano22}, we have not comprehensively matched our catalogue with all previous X-ray or optical studies of our sample. This will be the subject of future papers.

\subsubsection{Statistical Excess}

Here we show that there is evidence for a statistical excess of radio continuum sources associated with a subset of the clusters in our sample, although it will turn out that this is largely due to the individual sources already identified above. We make this statistical calculation for two different regions of interest in the cluster: the core radius (within which dynamically segregated compact objects are preferentially found; \citealt{1987IAUS..125..187V}), and the half-light radius (relevant for an unsegregated cluster radio source population).

First, we count the number of independent sources detected at 7.25 GHz within the core and half-light radius for each cluster, limiting the count to those sources bright enough to have been detected out to the edge of the core or half-light radius respectively. 
We then use our mean background source counts (as calculated in Section 4.1) to estimate the expected number of background sources down to the relevant flux density limit. We chose this approach rather than a local background calculation, since the lower sensitivity in the outer region of the primary beam means that the per-cluster background is very noisy. One potential concern for this approach is that the surface density of Galactic  sources is expected to depend on Galactic latitude, but the analysis of the MAVERIC VLA sample \citep{Shishkovsky20} implies that Galactic foreground/background sources have a negligible effect on these calculations.

The results of this analysis for the core radii are shown in Table~\ref{tab:n_excess_indi},
which for each cluster lists the number of sources above the flux density limit within each core, the expected number of background sources, and the Poissonian probability that the observed population of sources within the core radius is due entirely to the background. 

There are seven clusters for which the core population of radio sources is significantly ($\sim2$--$3\sigma$) above that expected from the background: Terzan 5, Terzan 1, Terzan 6, NGC\,6624, NGC\,6352, NGC\,6652, and NGC\,6388. For Terzan 5, Terzan 6, NGC\,6624, and NGC\,6652, the detected sources are already known and were discussed above. Terzan 1 is an obscured bulge cluster with several X-ray sources, including the historical X-ray burster XB 1732--304 \citep{1981ApJ...247L..23M,2006MNRAS.369..407C}, and the radio source could be the counterpart to this source or to a different one. NGC\,6388 also has a high predicted collision rate, and contains a substantial population of X-ray sources \citep{2012ApJ...756..147M} and likely pulsars, though the latter have not yet been detected. NGC\,6352 is an outlier, with a larger core than these other clusters and a relatively low collision rate; it may be that it simply is a statistical outlier in the number of background sources, which would not be too surprising given the $\sim 2\sigma$ significance of its excess and our sample size (26 clusters).

As discussed above, there are confirmed cluster sources even in clusters that do not formally show a statistically significant excess, such as 47 Tuc and NGC\,6397 (albeit outside the small core radius for this core-collapsed cluster). Hence other clusters without evidence for such an excess may still contain true cluster member sources in our catalogue. It is clear that 
due to the relatively high density of background sources, an excess is more readily detected in those clusters with very small core radii.

Unlike the case for the core radii calculation, we do not tabulate the results for the half-light radii, since only two clusters show a statistically significant excess: NGC\,6652 ($N_{r_h} = 5$, $N_{\rm bkg, r_h} = 0.9$, $p=0.002$) and Terzan 5 ($N_{r_h} = 3$, $N_{{\rm bkg}, r_h} = 0.6$, $p=0.02$). It seems likely that the majority of the 7.25 GHz radio continuum sources within the half-light radii of our sample clusters are indeed background sources.

\begin{table}
\centering
\caption{Counts of sources within the cluster core radii at 7.25 GHz. We report the number of detected sources ($N$), the expected number of background sources ($N_{\rm bkg}$), and the probability that a core source population is consistent with the expected background (no value is listed for clusters without core sources). The table is sorted from most significant detection to least significant.}
\label{tab:n_excess_indi}
\begin{tabular}{lrrl}
\hline \hline
      GC &  N$_{\rm core}$ &  N$_{\rm bkg}$ & $p$   \\
\hline
Terzan 5   & 2 & 0.03 &  0.001    \\
Terzan 1   & 1 & 0.00 &  0.004  \\
Terzan 6    & 1 & 0.01 &  0.006  \\
NGC\,6624 & 1 & 0.01 & 0.012    \\
NGC\,6352 & 6 & 2.2 &  0.017   \\
NGC\,6652 & 1 & 0.04 & 0.040   \\
NGC\,6388 & 1 & 0.06 &  0.054  \\
47 Tuc      & 1 & 0.35 & 0.24   \\
NGC\,4372    & 5 & 6.9 & 0.34    \\
NGC\,6362   & 3 & 4.6 &  0.40   \\
$\omega$ Cen & 4 & 10.3 &  0.48    \\
NGC\,4833 & 1 & 2.8 & 0.52   \\
NGC\,3201 & 1 & 3.4 &  0.54   \\
Djorg 2  & 0 & 0.30 &  ---  \\
Liller1  & 0 & 0.01 &  ---  \\
M54      & 0 & 0.01 & ---  \\
NGC\,2808 & 0 & 0.20 &  ---  \\
NGC\,5927 & 0 & 0.45 &  ---  \\
NGC\,6139 & 0 & 0.06 &  ---  \\
NGC\,6304 & 0 & 0.14 &   ---  \\
NGC\,6397 & 0 & 0.01 &   ---  \\
NGC\,6441 & 0 & 0.06 &   ---  \\
NGC\,6522 & 0 & 0.01 &  ---  \\
NGC\,6541 & 0 & 0.07 &   ---  \\
NGC\,6553 & 0 & 0.83 &   ---  \\
NGC\,6752 & 0 & 0.09 &  ---    \\
\hline
\end{tabular}
\end{table}

\section{Discussion and conclusions}
\label{sec:discuss}

Out of the 1285 radio continuum sources in our catalogue, a minority appear to be associated with the globular clusters in the target field. Seven of the 26 clusters show a statistically-significant excess of sources within their core radii, while only two of 26 show such an excess within their half-light radii. Overall, the implication is that globular clusters with radio continuum sources persistent at $S_\nu \gtrsim 20$\,\si{\micro}Jy at 7 GHz are the exception rather than the rule. There is strong evidence that this conclusion is sensitivity-limited: deeper data from the VLA have revealed fainter X-ray binaries in some globular clusters, such as the black hole candidate in M62 \citep{2013ApJ...777...69C}. Hence, deeper studies with next-generation radio facilities are required to sample this population to higher completeness. The Square Kilometre Array (SKA) precursor facility MeerKAT \citep{Jonas16} can provide complementary coverage of Southern globular clusters at lower frequencies than presented here, with consequently higher sensitivity to steep-spectrum objects such as pulsars. However, its angular resolution (currently $\sim 6$\arcsec{} at 0.9--1.6 GHz) makes it less well suited to detailed imaging studies of the most compact or crowded globular clusters, and implies that source confusion limits the currently achievable noise levels\footnote{\url{https://apps.sarao.ac.za/calculators/continuum}} to $\sim 3$\,\si{\micro}Jy\,bm$^{-1}$. Both resolution and imaging depth will improve when the new S-band (1.6--3.5 GHz) receivers are fully installed on MeerKAT, enabling somewhat deeper images than presented in the current work. Towards the end of the decade, the longer baselines, higher frequencies and higher sensitivity of SKA1-Mid will enable a new generation of southern-hemisphere globular cluster imaging studies.

Despite the limited statistical signal in many of the sample clusters, our ATCA survey has already been successful in identifying new compact binaries in globular clusters, such as the ultracompact black hole X-ray binary candidate X9 in 47 Tuc \citep{2015MNRAS.453.3918M, 2017MNRAS.467.2199B} and the transitional millisecond pulsar candidate NGC6652B \citep{Paduano21}. This highlights the need and opportunity for detailed follow-up of each central radio source in our catalogue---many are likely to be true cluster sources, even if the individual cluster does not formally show a statistical excess of sources.

\citet{2020ApJ...898..162W} identify a number of clusters that, on the basis of their low degree of mass segregation, are theoretically consistent with having a large population of stellar-mass black holes. Of these clusters, two are in the present sample: NGC\,2808 and NGC\,5927. Neither of these has evidence for an excess of radio sources, though there is not necessarily a correlation between a substantial population of black holes and the presence of ones in close enough binaries to undergo mass transfer \citep{2018ApJ...852...29K}, and both of these clusters are distant ($\gtrsim 8$ kpc). We also note that NGC\,6388, for which we found a core radio source excess, is one of the clusters that has been singled out by \mbox{\citet{2009ApJ...690.1370M}}, based on its high mass and metallicity, as potentially retaining a large population of black holes. NGC\,6441 has similar properties. Neither of these clusters were included in the \citet{2020ApJ...898..162W} study, but are promising clusters deserving of follow-up efforts.

Although our survey was designed to be sensitive to any intermediate-mass black holes located close to the cluster centres, our data show no evidence for any radio emission associated with such objects, which would arise from the jets associated with low-level accretion from the cluster gas. Subject to a plausible set of assumptions, \citet{2018ApJ...862...16T} placed stringent upper limits of 610--2270\,$M_{\odot}$ on the masses of any intermediate-mass black holes at the centres of our clusters.  Should they exist, deeper radio and X-ray observations would be needed to detect their accretion signatures.

This work shows that modern radio continuum surveys are deep enough to find at least some radio sources associated with Galactic globular clusters, especially in the nearer clusters. However, the nature of most of these sources is difficult to ascertain from radio observations alone. To reach scientific goals such as identifying candidate stellar-mass black holes, multi-wavelength observations are needed, as evidenced by the discovery and classification 
of the black hole candidates 47\,Tuc\,X9 \citep{2015MNRAS.453.3918M} and M62-VLA1 \citep{2013ApJ...777...69C}. 

Current radio surveys are not yet deep enough to access typical quiescent black hole X-ray binary radio flux densities. At a few kpc, sensitivities reach $L_{\rm R} \approx 10^{28}$\,erg\,s$^{-1}$ in radio luminosity, corresponding to $L_{\rm X} \approx 10^{33}$\,erg\,s$^{-1}$ on the main $L_{\rm R}-L_{\rm X}$ correlation \citep[e.g.][]{2014MNRAS.445..290G}. This X-ray luminosity is at the top end of typical quiescent luminosities \citep{2008ApJ...683L..51G}, appropriate either for long orbital period systems ($\sim 10$\,hr), or very short orbital period systems ($\lesssim 1$\,hr) such as ultracompacts, but which excludes the fainter $L_{\rm X}$ expected from more typical main sequence secondaries.
There are suggestive clues that some black holes may appear more radio-loud than others at a similar X-ray luminosity, such as the two candidates in M22 \citep{2012Natur.490...71S} and in the foreground of M15 \citep{2016ApJ...825...10T}, but 
ultimately dynamical mass measurements at optical or IR wavelengths are still necessary to answer the most pressing questions about the occurrence and properties of compact objects in globular clusters. This work is an initial step in this exciting long-term effort that shows radio continuum observations are an important part of the overall multi-wavelength picture.

\section*{Acknowledgements}

VT acknowledges a CSIRS scholarship from Curtin University. JCAM-J is the recipient of an Australian Research Council Future Fellowship (FT140101082). RMP acknowledges support from Curtin University through the Peter Curran Memorial Fellowship. COH and GRS are funded by NSERC Discovery Grants RGPIN-2016-04602 and  RGPIN-06569-2016 respectively. JS acknowledges support from NSF grants AST-1308124 and AST-1514763, NASA grants 80NSSC21K0628, Chandra-GO8-19122X, and HST-GO-14203.002, and the Packard Foundation. TDR acknowledges financial contribution from ASI-INAF n.2017-14-H.0, an INAF main stream grant. 

The Australia Telescope Compact Array is part of the Australia Telescope National Facility (grid.421683.a) which is funded by the Australian Government for operation as a National Facility managed by CSIRO. We acknowledge the Gomeroi people as the traditional owners of the Observatory site. The International Centre for Radio Astronomy Research is a joint venture between Curtin University and the University of Western Australia, funded by the state government of Western Australia and the joint venture partners.

This research has made use of NASA's Astrophysics Data System, arXiv, SAOImage {\sc ds9} \citep{2003ASPC..295..489J}, and the Python libraries SciPy \citep{scipy}, NumPy \citep{van2011numpy}, AstroPy \citep{2013A&A...558A..33A}, Pandas \citep{mckinney-proc-scipy-2010} and Matplotlib \citep{Hunter:2007}.

This is a pre-copyedited, author-produced PDF of an article accepted for publication in Monthly Notices of the Royal Astronomical Society following peer review, published by Oxford University Press on behalf of the Royal Astronomical Society. The version of record is available online at \url{https://academic.oup.com/mnras/advance-article-abstract/doi/10.1093/mnras/stac1034/6568559}, or at \url{https://www.doi.org/10.1093/mnras/stac1034}.


\section*{Data Availability}
The original data underlying this article may be accessed from the Australia Telescope Online Archive (\url{https://atoa.atnf.csiro.au/query.jsp}), under project codes C2158 and C2877. The full source catalogue is available in the article and in its supplementary material. The final FITS images of all globular cluster fields are available from the CSIRO Data Access Portal, at \url{https://data.csiro.au/collection/csiro:54270} \citep{maveric-s}. 


\bibliographystyle{mnras}
\input{maveric.bbl}



\appendix

\section{Cluster images}
In Figure~\ref{fig:allradio} we provide 5.5-GHz images of all 26 clusters covered in this study. We indicate the detected sources, as well as the core and half-light radius of each cluster.

\begin{figure*}
\begin{subfigure}[t]{\hsize}
\centering
\includegraphics[width=8.5cm]{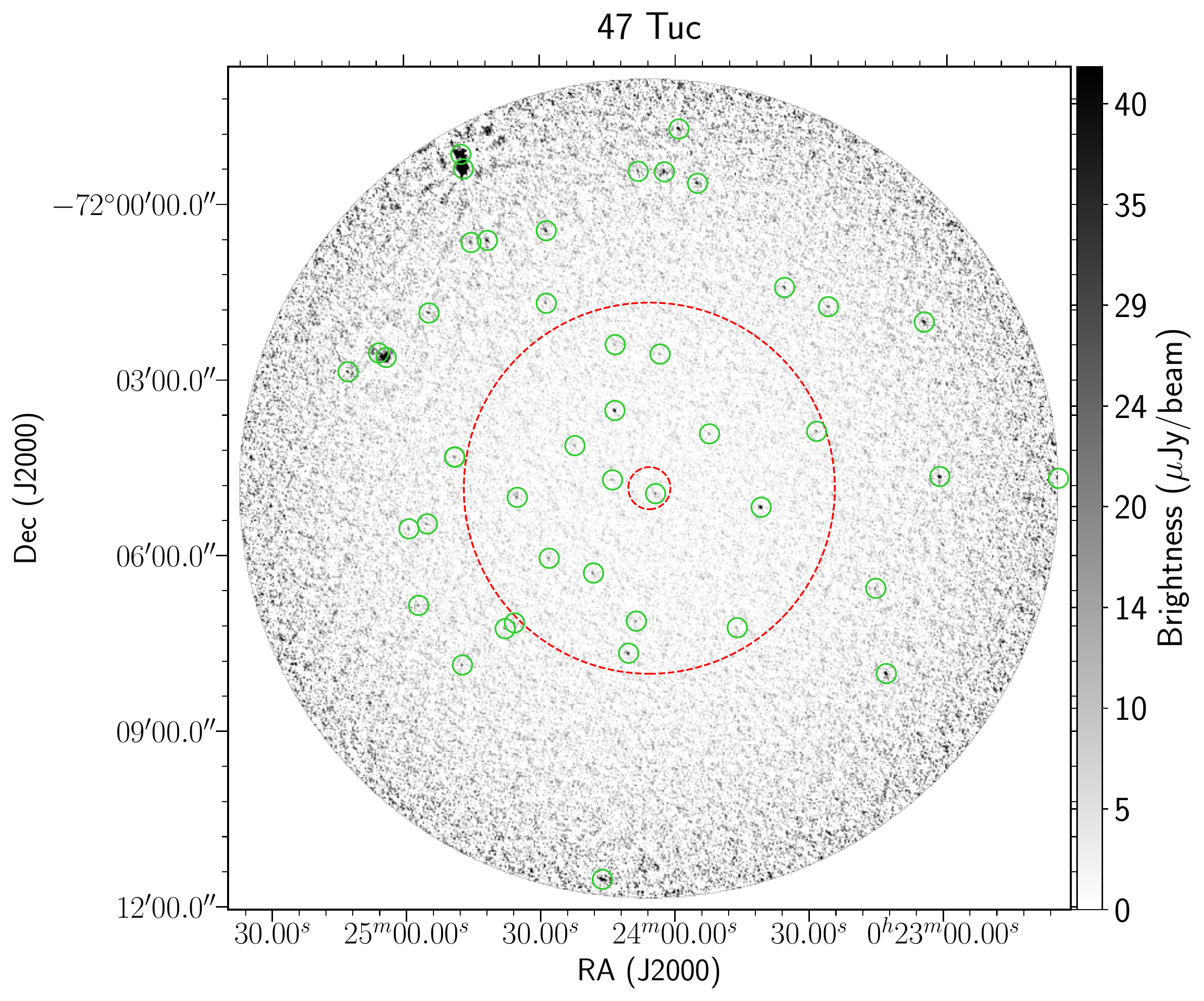}
\includegraphics[width=8.5cm]{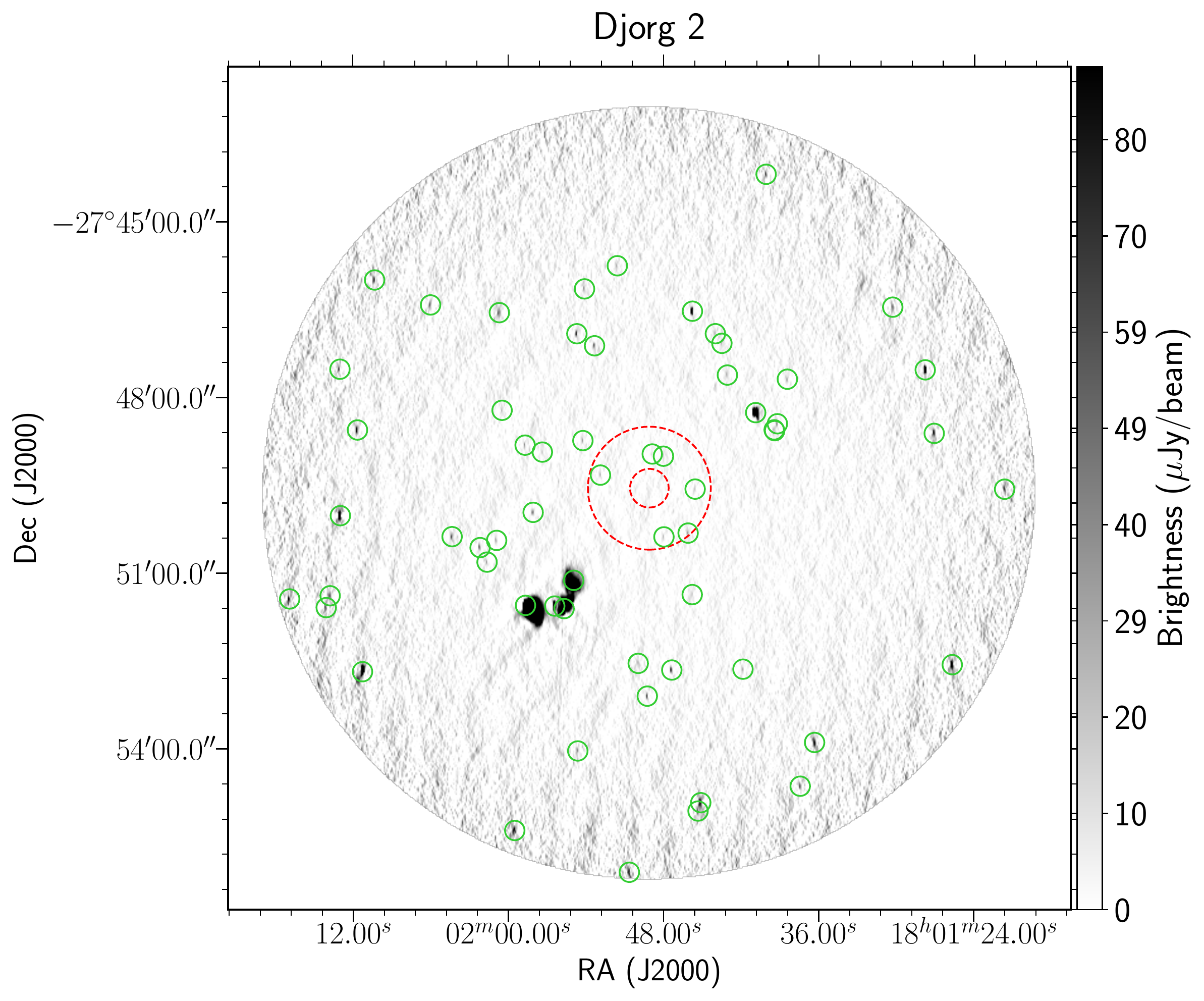}
\end{subfigure}   
\begin{subfigure}{\hsize}
\centering
\includegraphics[width=8.5cm]{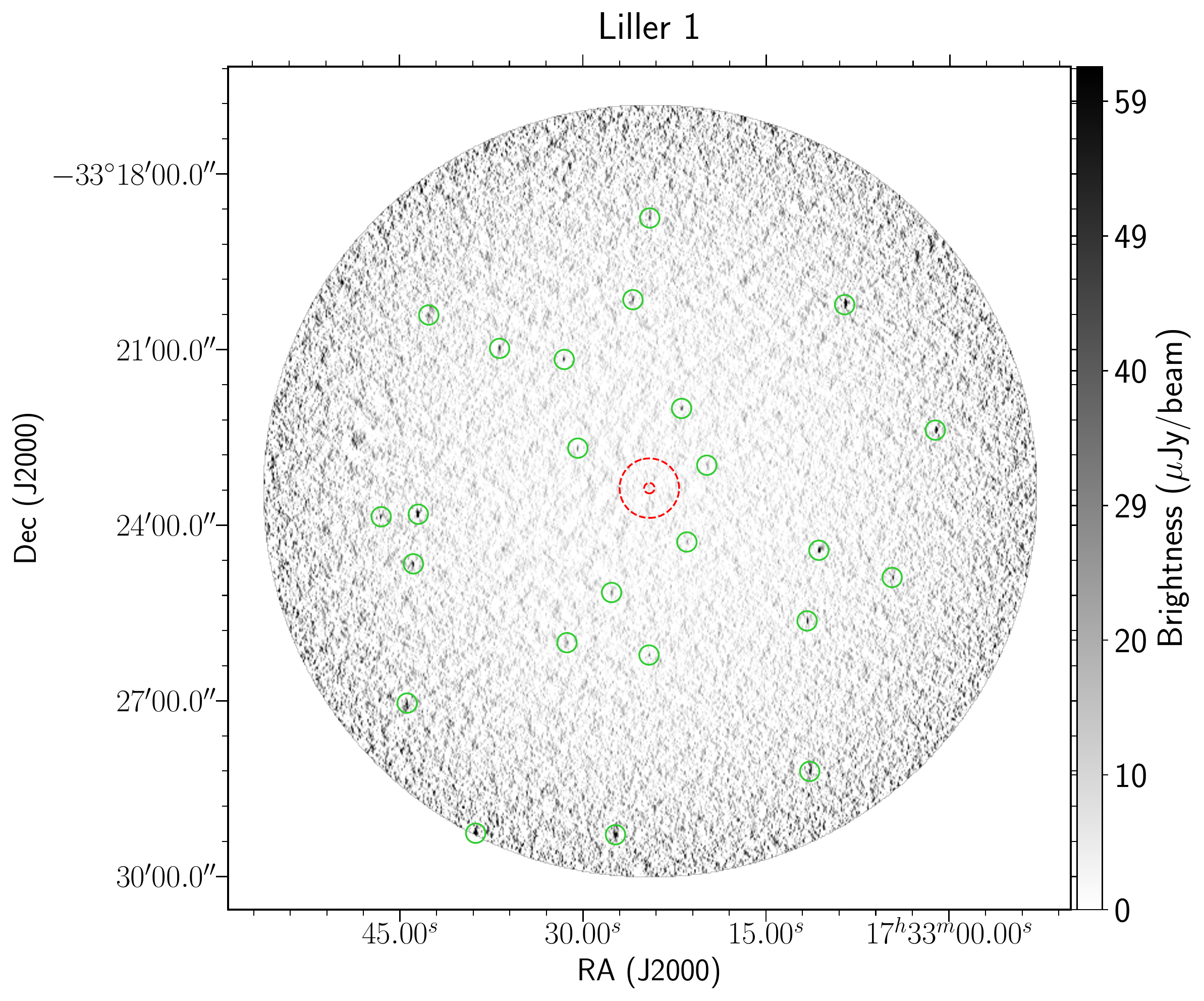}
\includegraphics[width=8.5cm]{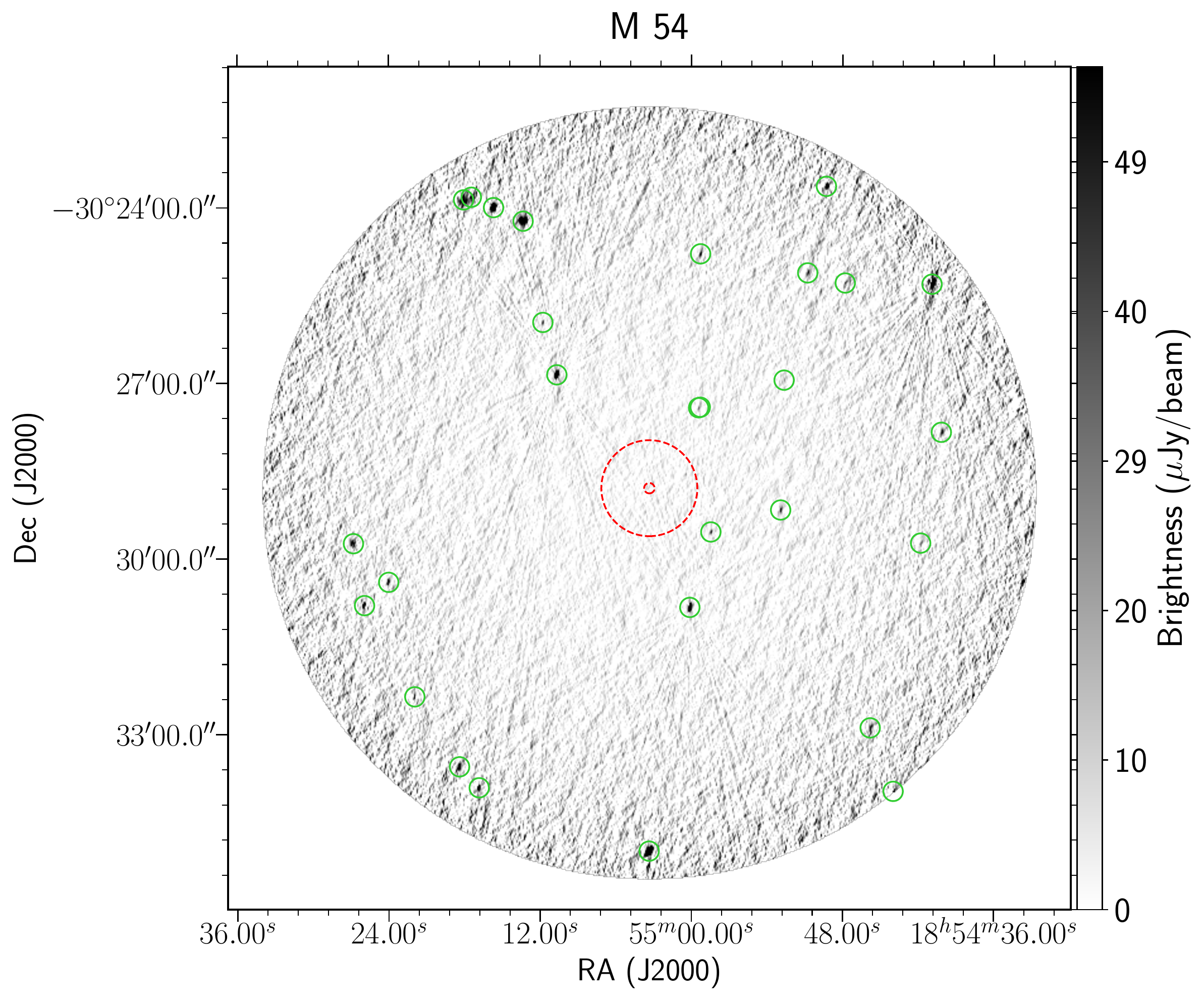}
\end{subfigure}   
\begin{subfigure}{\hsize}
\centering
\includegraphics[width=8.5cm]{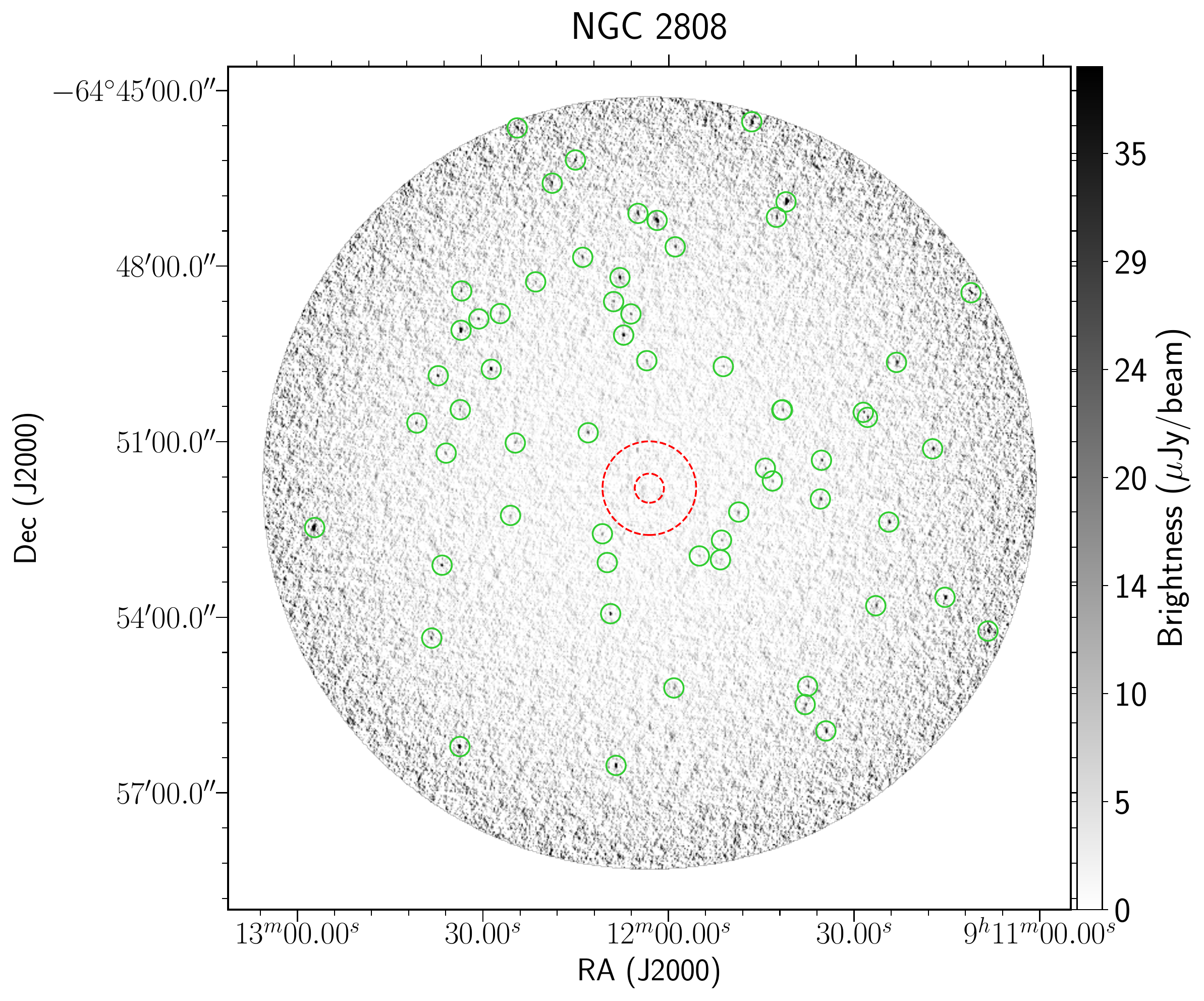}
\includegraphics[width=8.5cm]{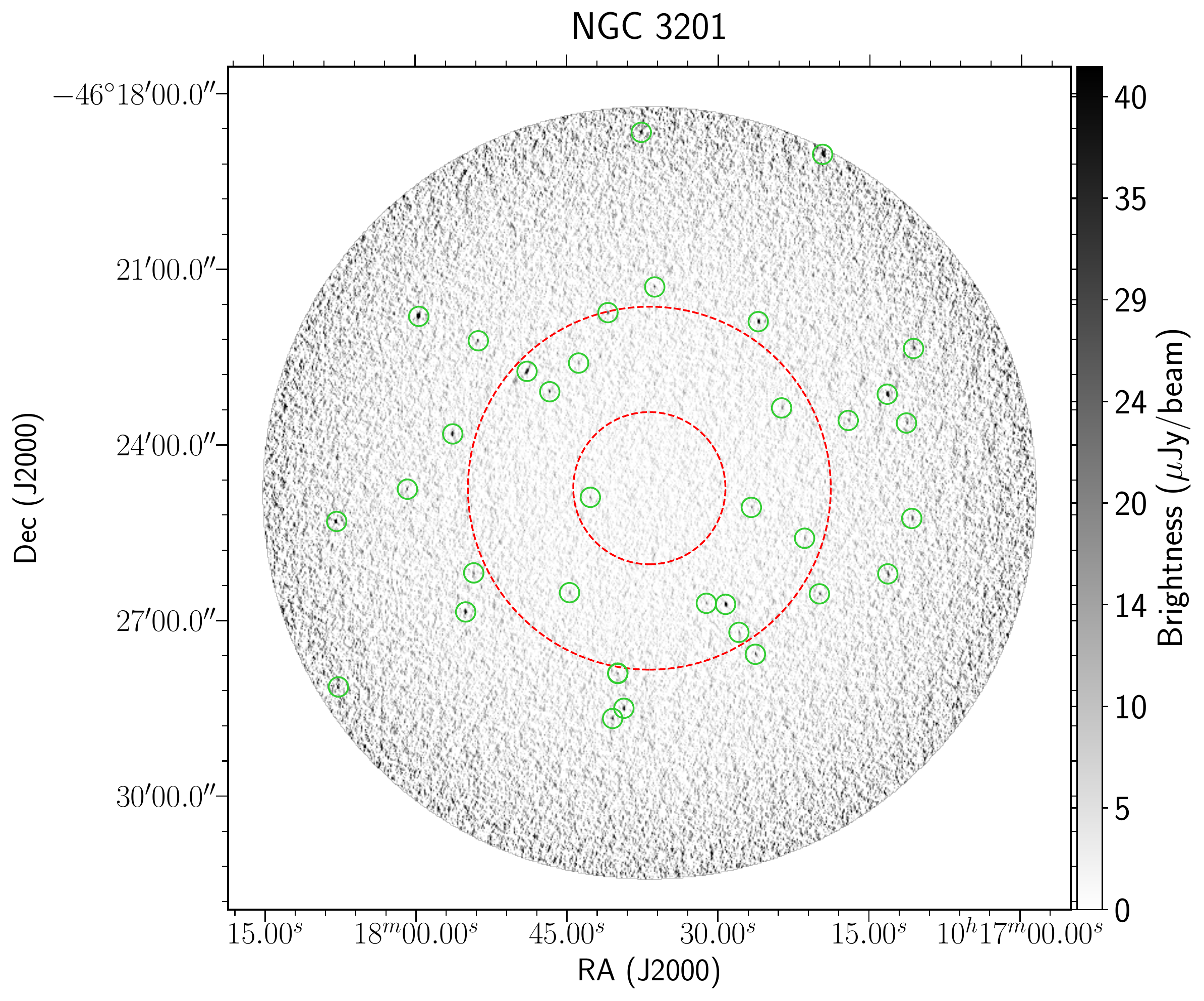}
\end{subfigure}
\caption{5.5-GHz ATCA images of the fields of 47 Tuc, Djorg 2, Liller 1, M54, NGC\,2808, and NGC\,3201. The core and half-light radii for each cluster are shown by the inner and outer dashed red lines, respectively. Detected sources are highlighted with small green circles. Image brightness is indicated by the colourbar, scaled appropriately for each figure.}
\end{figure*}

\begin{figure*}\ContinuedFloat
\begin{subfigure}[t]{\hsize}
\centering
\includegraphics[width=8.5cm]{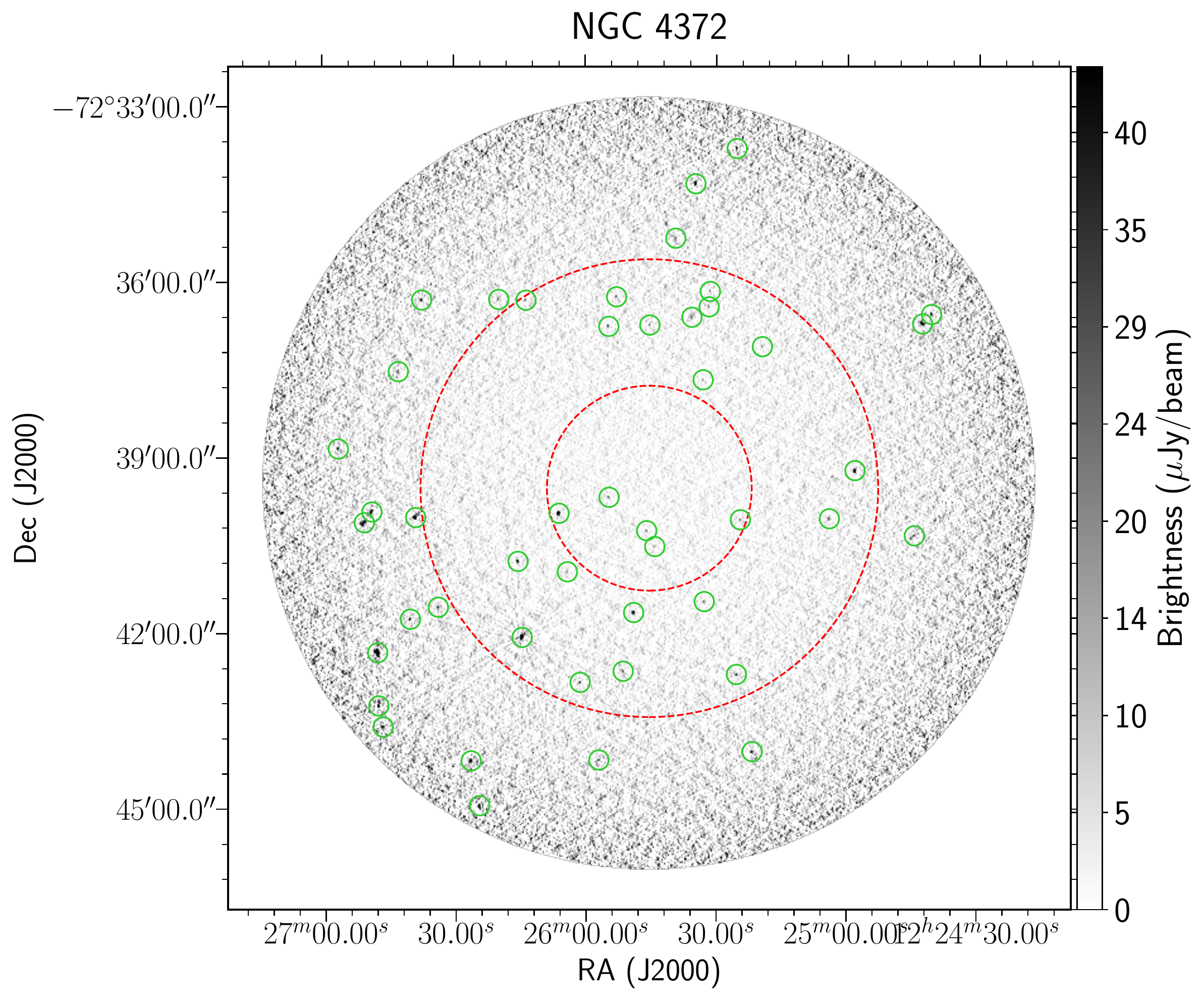}
\includegraphics[width=8.5cm]{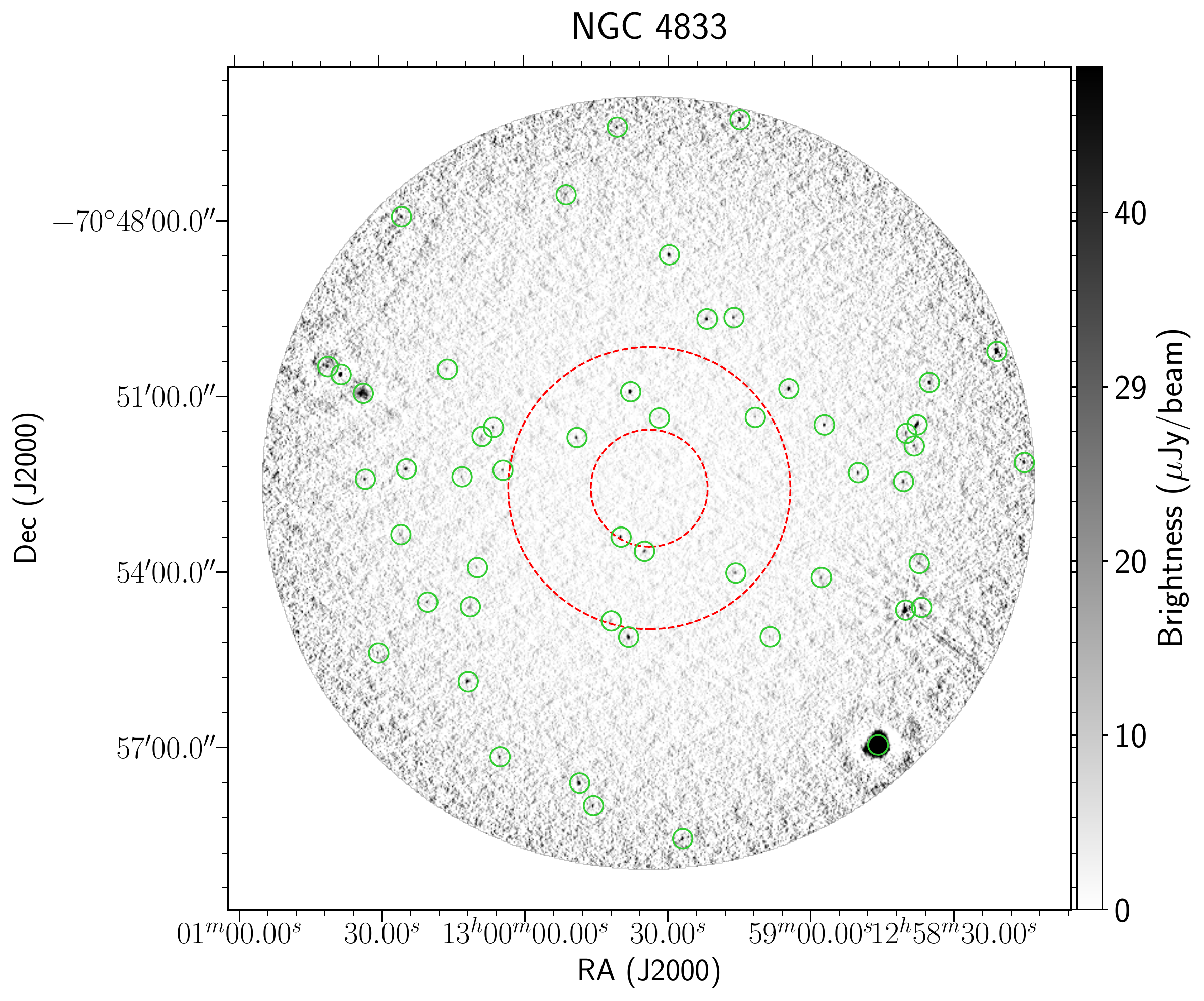}
\end{subfigure}   
\begin{subfigure}{\hsize}
\centering
\includegraphics[width=8.5cm]{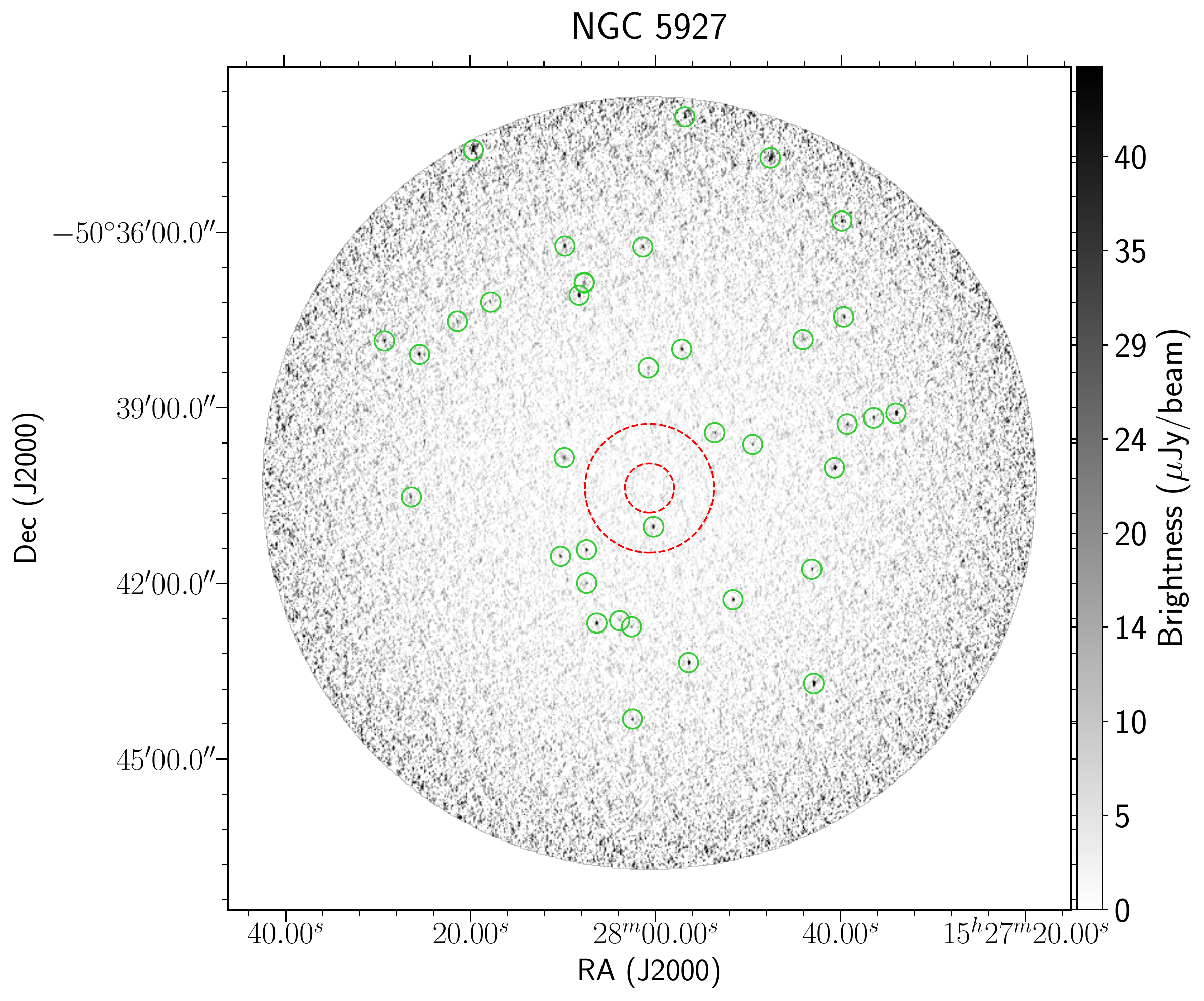}
\includegraphics[width=8.5cm]{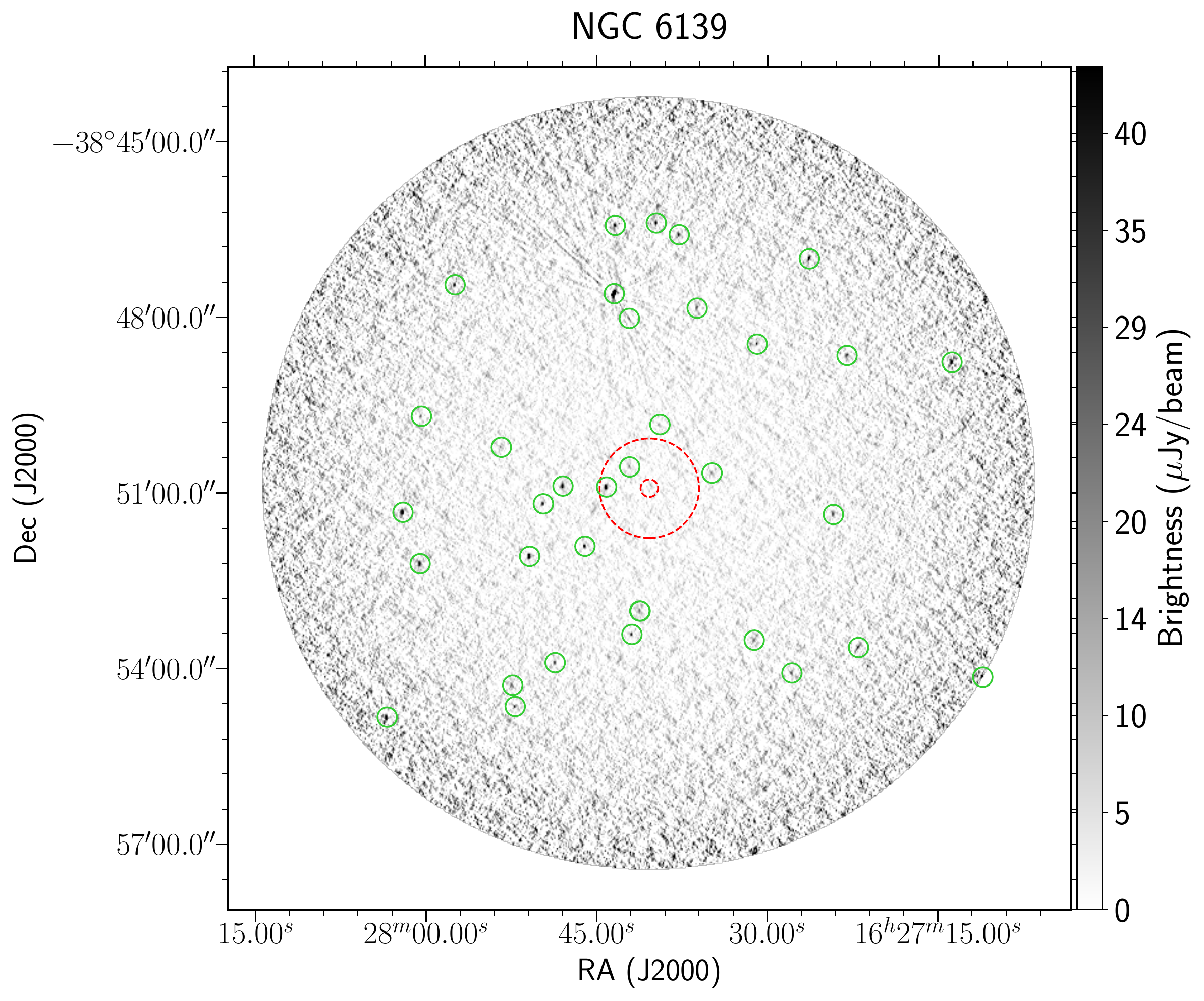}
\end{subfigure}   
\begin{subfigure}{\hsize}
\centering
\includegraphics[width=8.5cm]{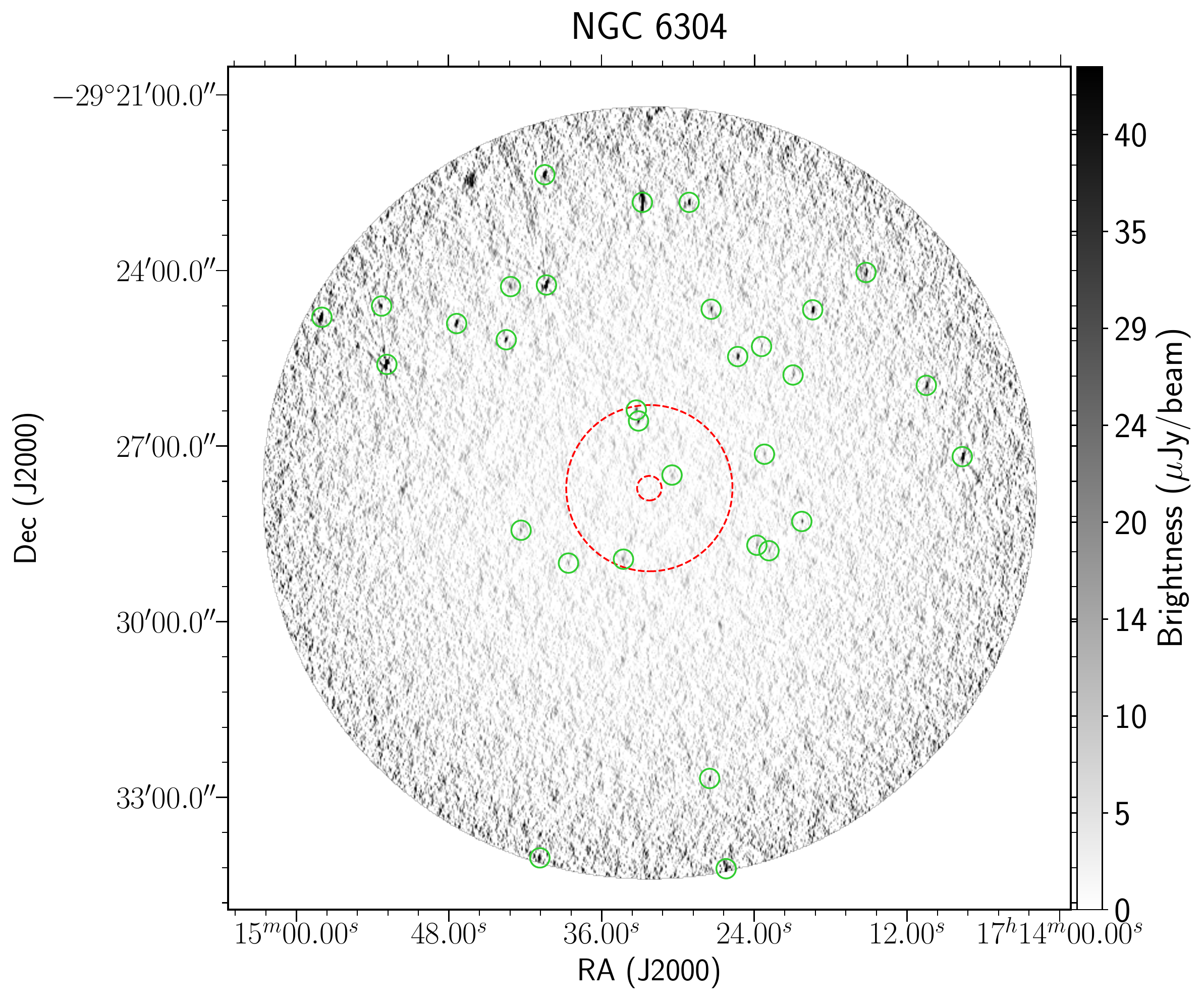}
\includegraphics[width=8.5cm]{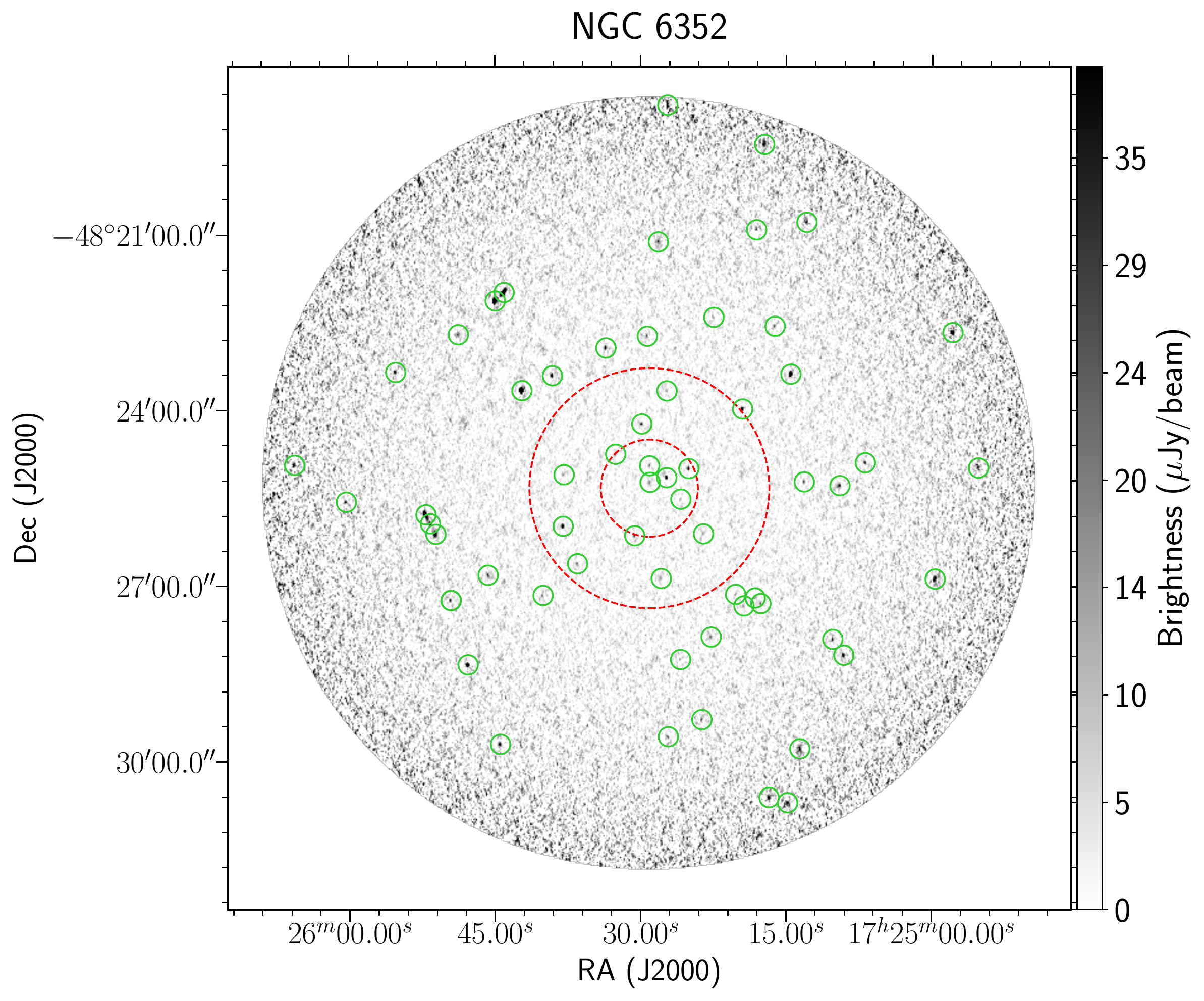}
\end{subfigure}   
\caption{5.5-GHz ATCA images of the fields of NGC\,4372, NGC\,4833, NGC\,5927, NGC\,6139, NGC\,6304 and NGC\,6352. The core and half-light radii for each cluster are shown by the inner and outer dashed red lines, respectively. Detected sources are highlighted with small green circles. Image brightness is indicated by the colourbar, scaled appropriately for each figure.}
\end{figure*}

\begin{figure*}\ContinuedFloat
\begin{subfigure}[t]{\hsize}
\centering
\includegraphics[width=8.5cm]{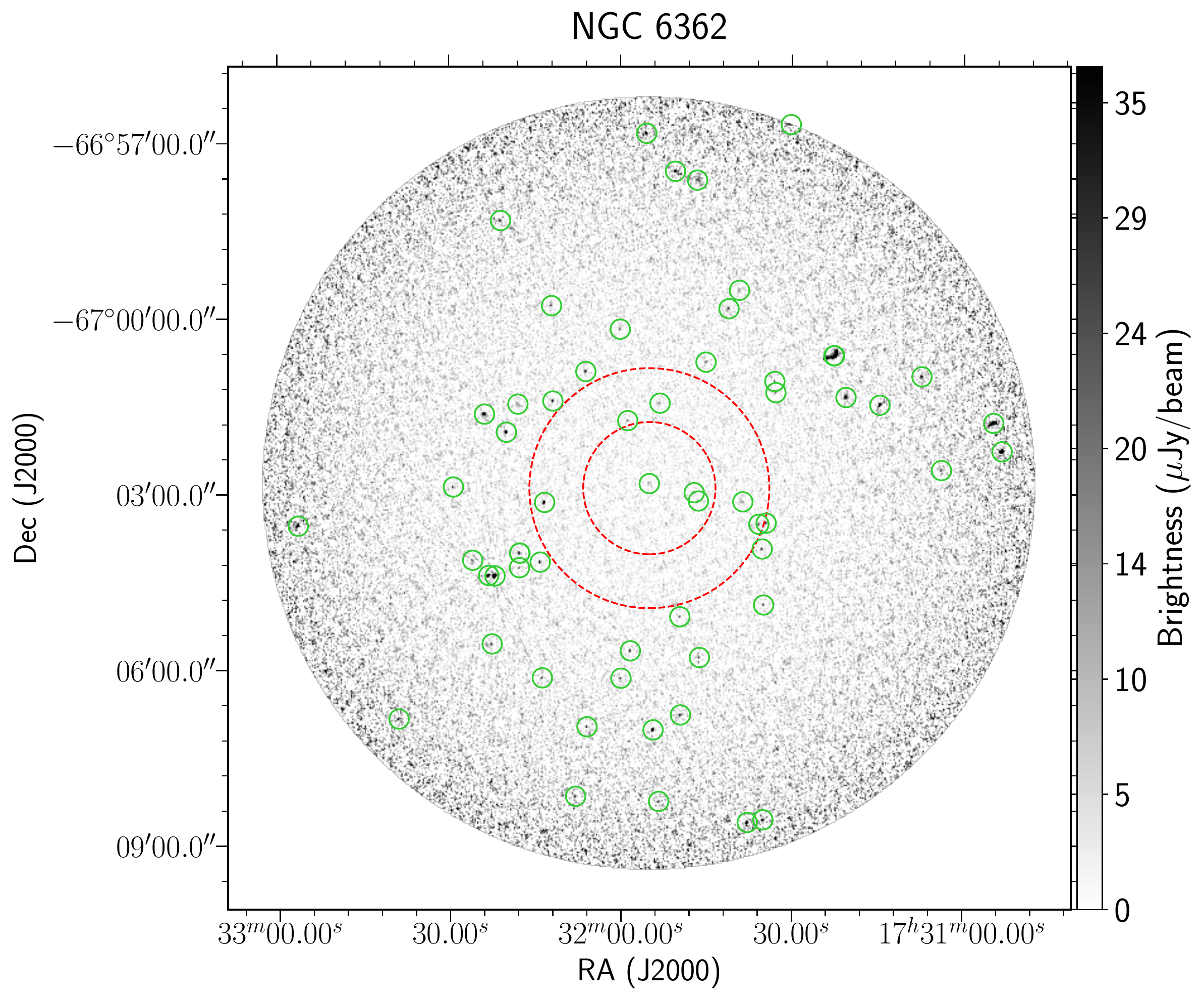}
\includegraphics[width=8.5cm]{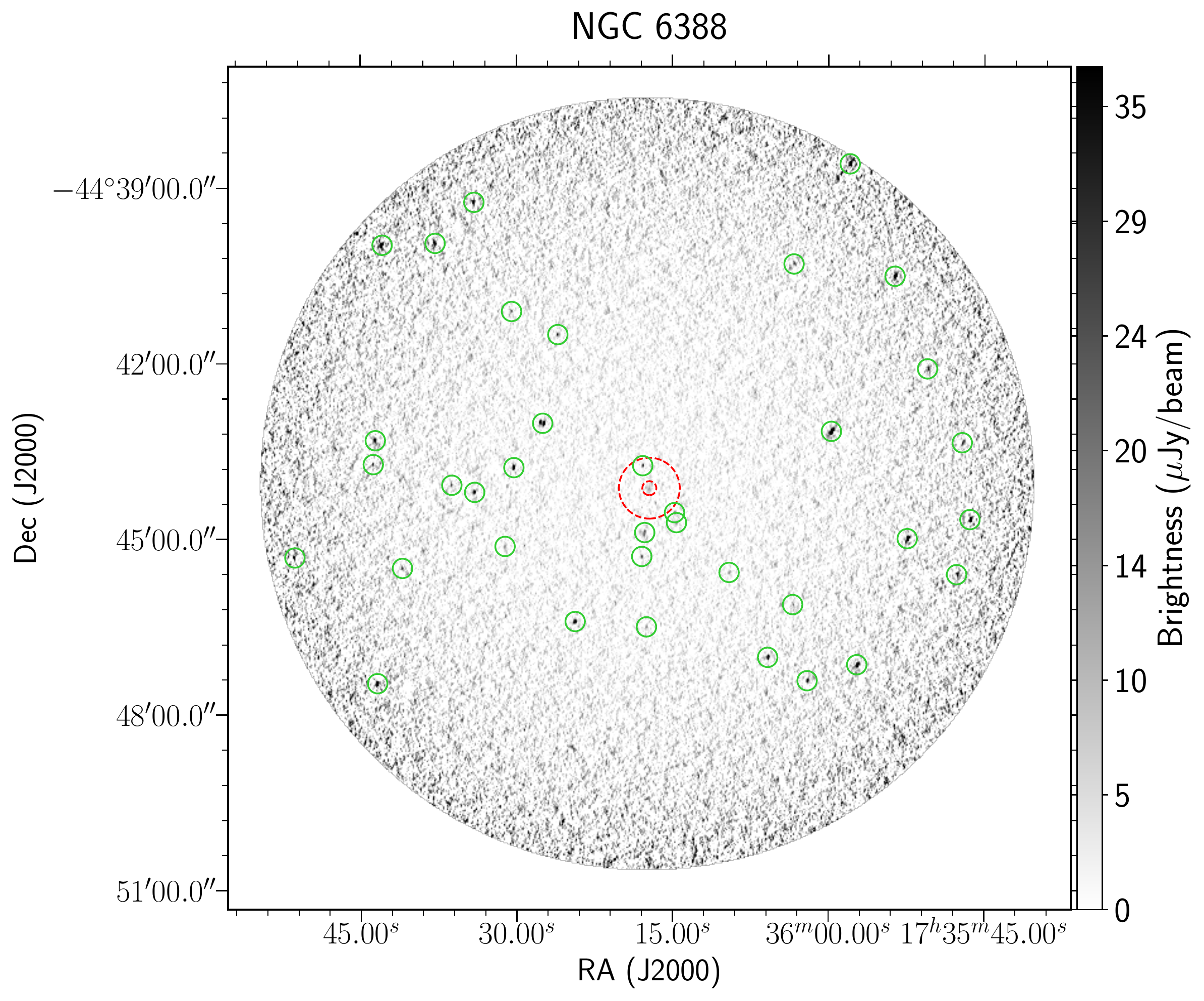}
\end{subfigure}   
\begin{subfigure}{\hsize}
\centering
\includegraphics[width=8.5cm]{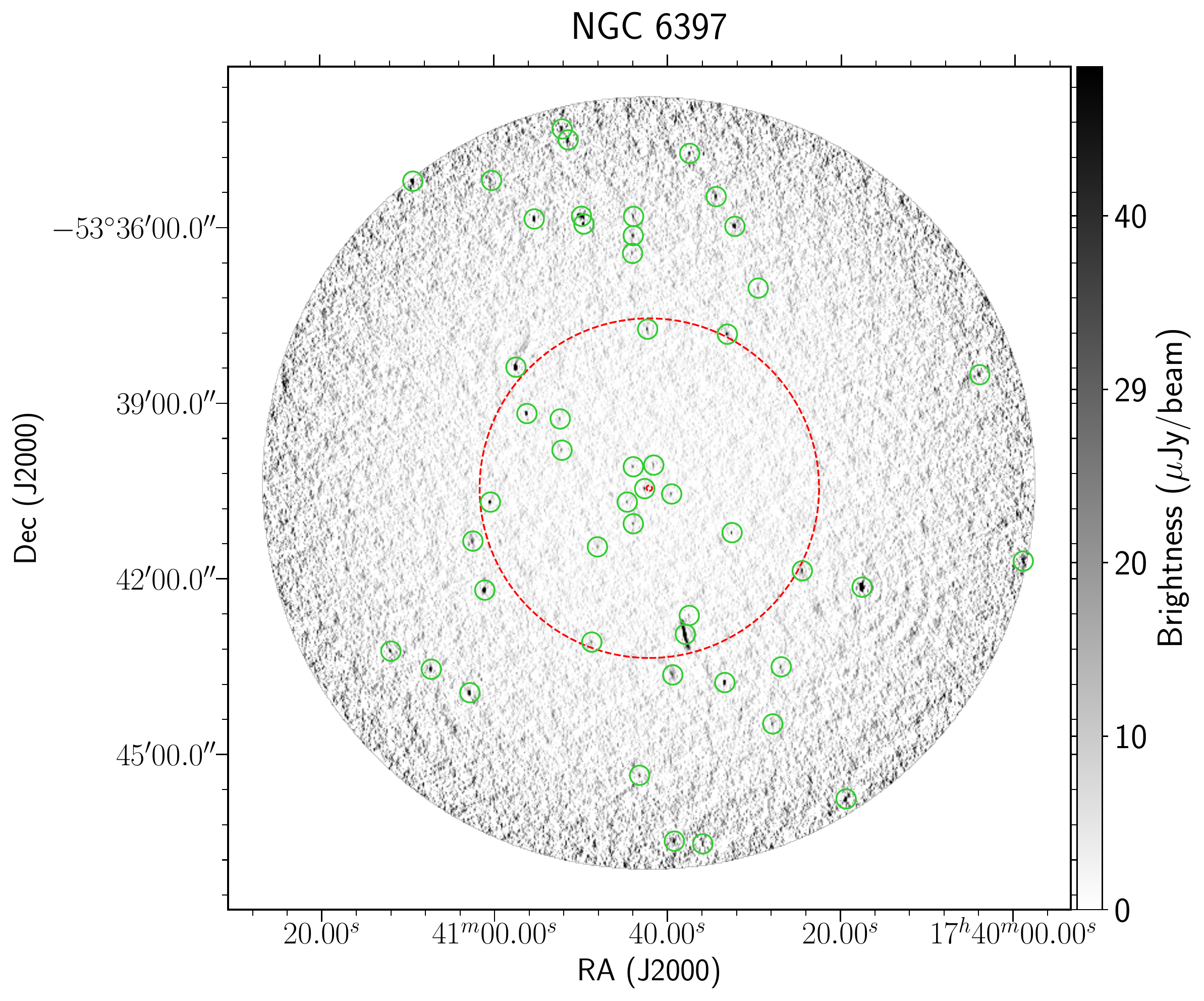}
\includegraphics[width=8.5cm]{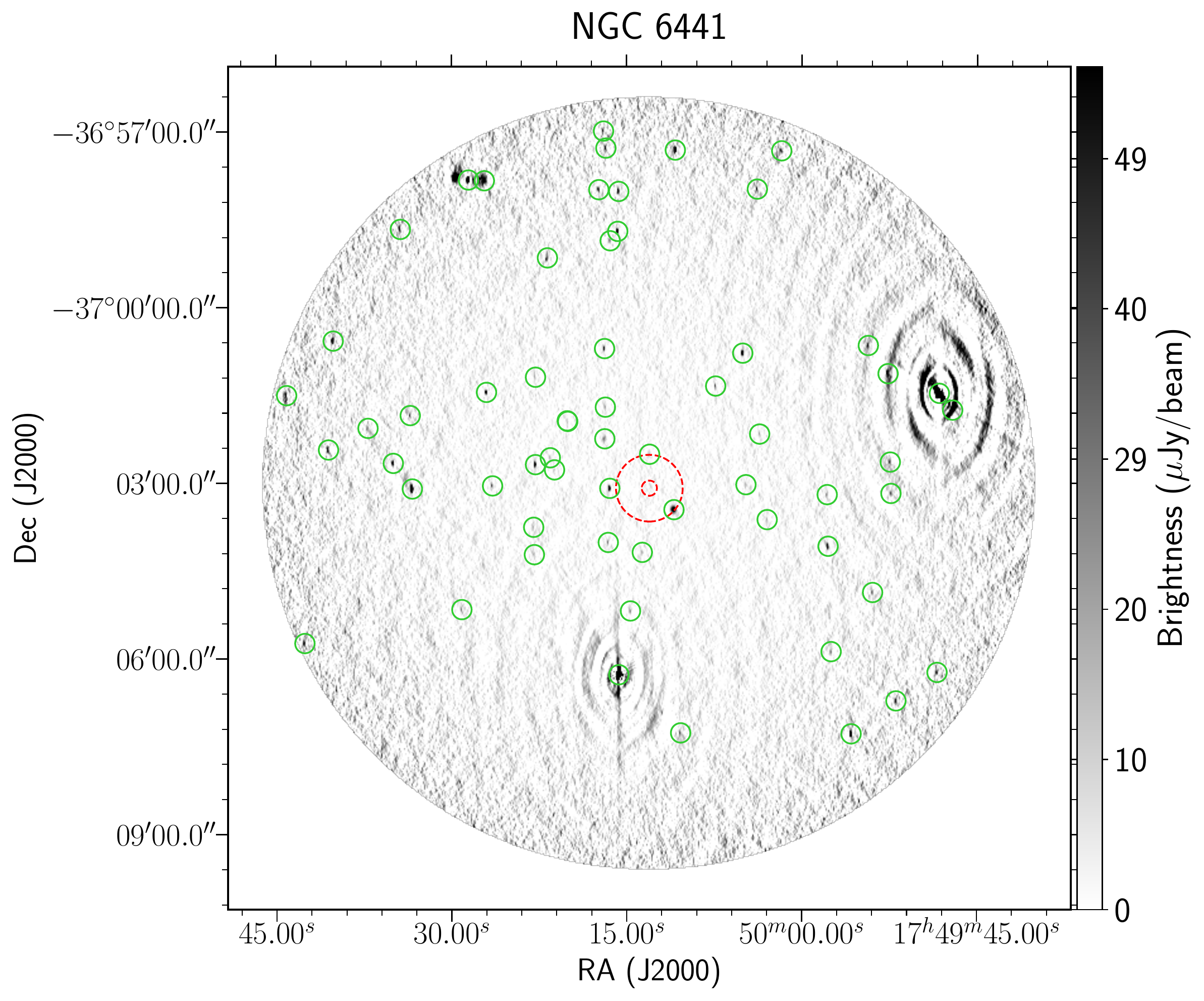}
\end{subfigure}   
\begin{subfigure}{\hsize}
\centering
\includegraphics[width=8.5cm]{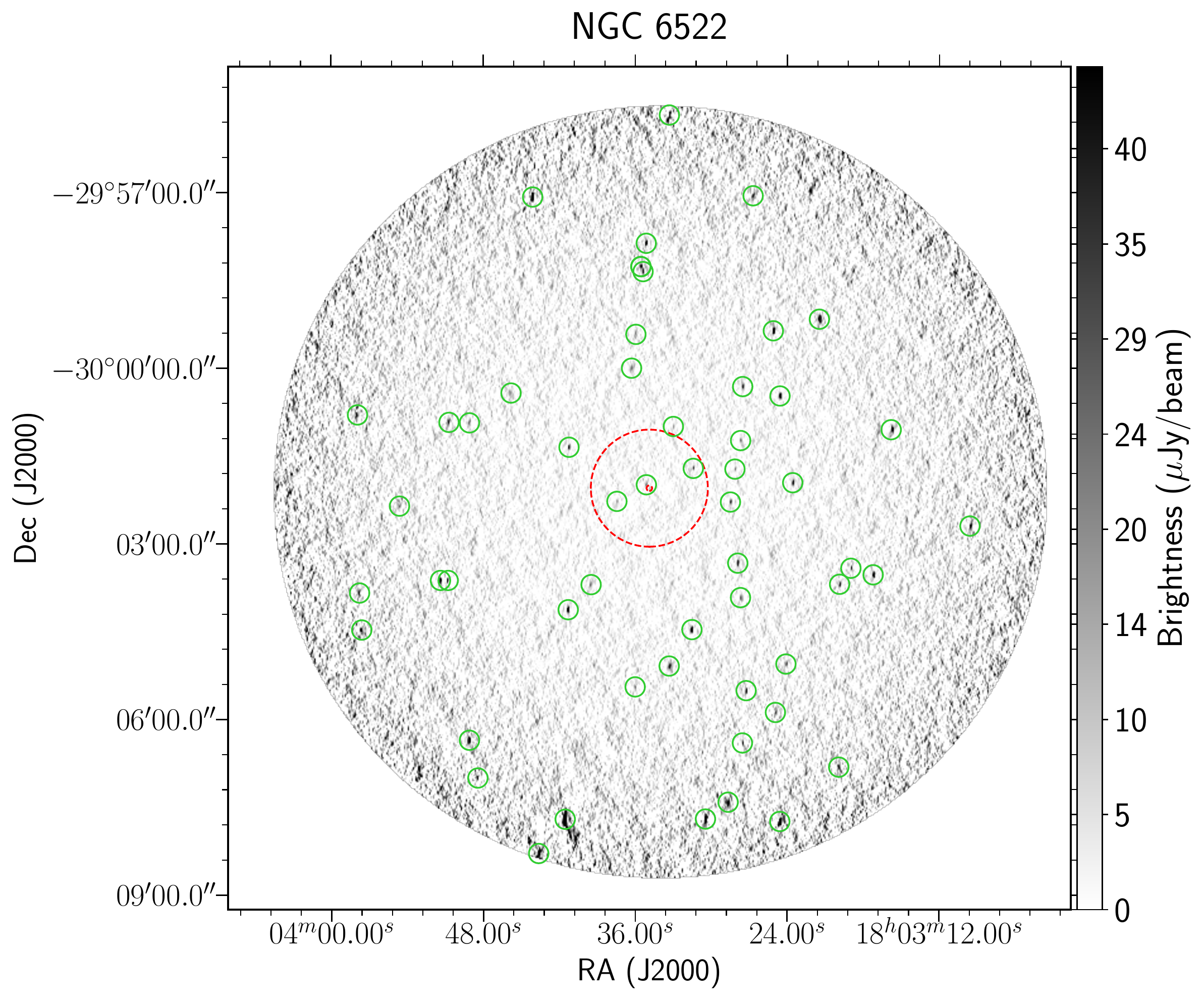}
\includegraphics[width=8.5cm]{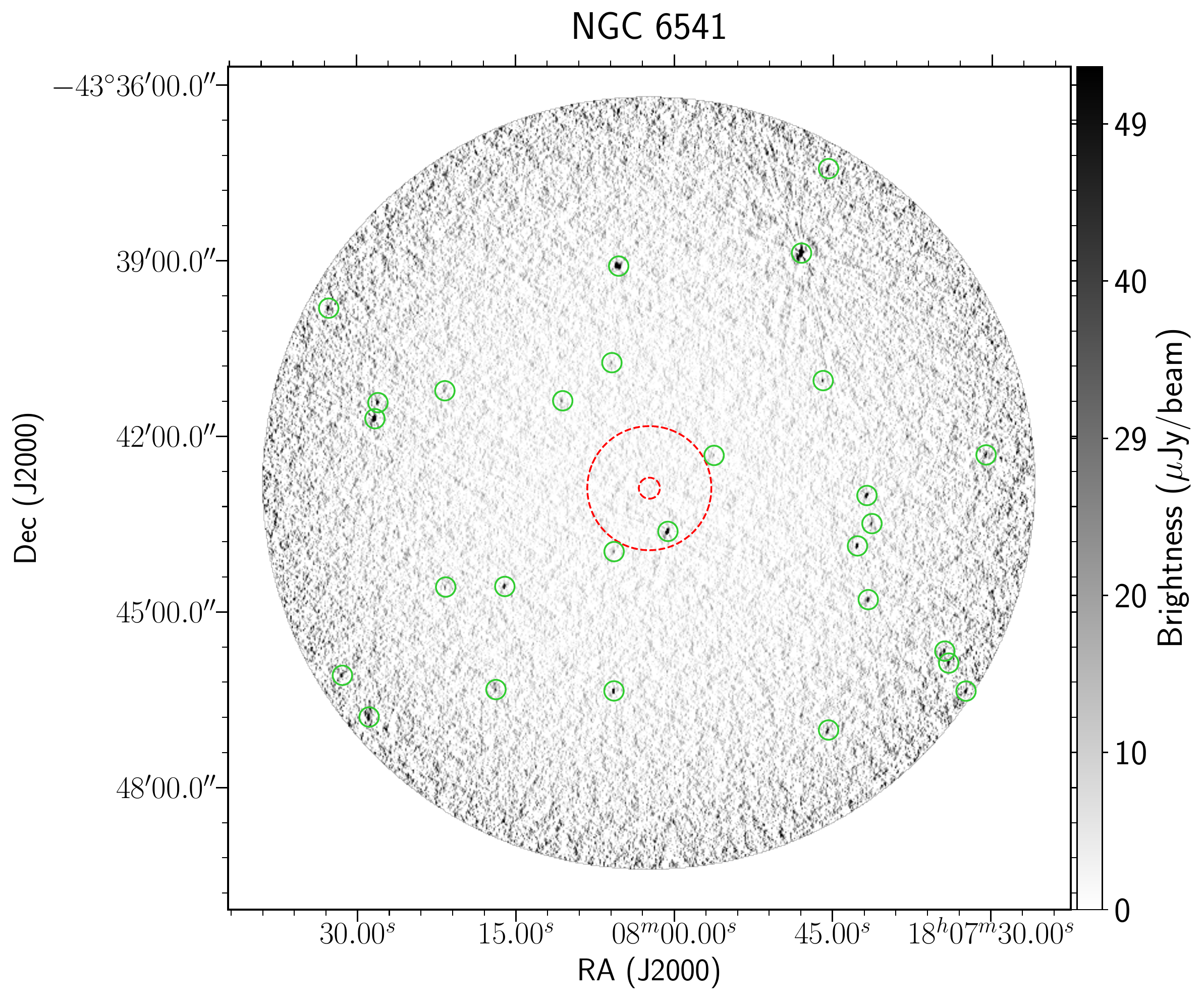}
\end{subfigure}   
\caption{5.5-GHz ATCA images of the fields of NGC\,6362, NGC\,6388, NGC\,6397, NGC\,6441, NGC\,6522, and NGC\,6541. The core and half-light radii for each cluster are shown by the inner and outer dashed red lines, respectively. Detected sources are highlighted with small green circles. Image brightness is indicated by the colourbar, scaled appropriately for each figure.}
\end{figure*}

\begin{figure*}\ContinuedFloat
\begin{subfigure}[t]{\hsize}
\centering
\includegraphics[width=8.5cm]{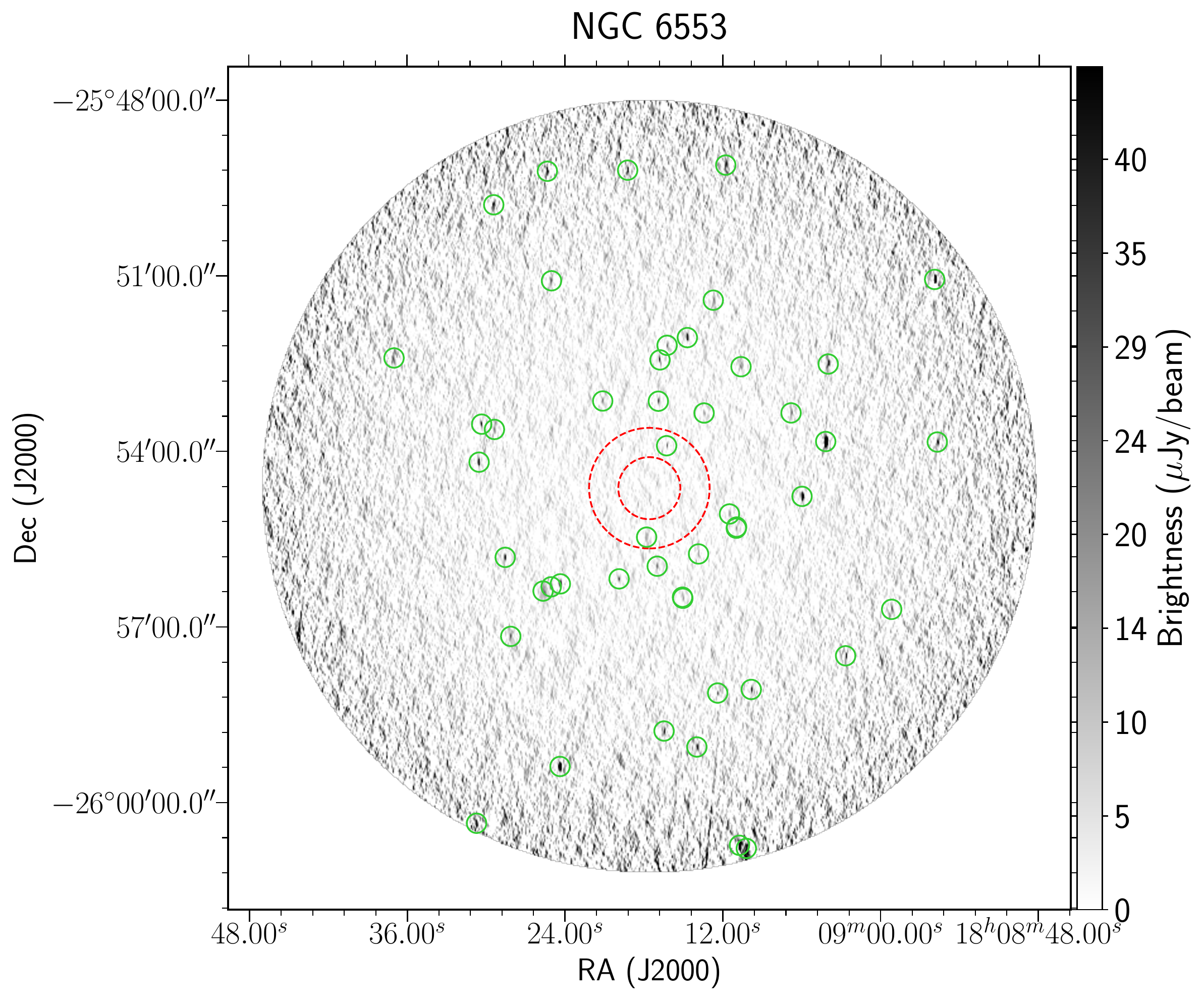}
\includegraphics[width=8.5cm]{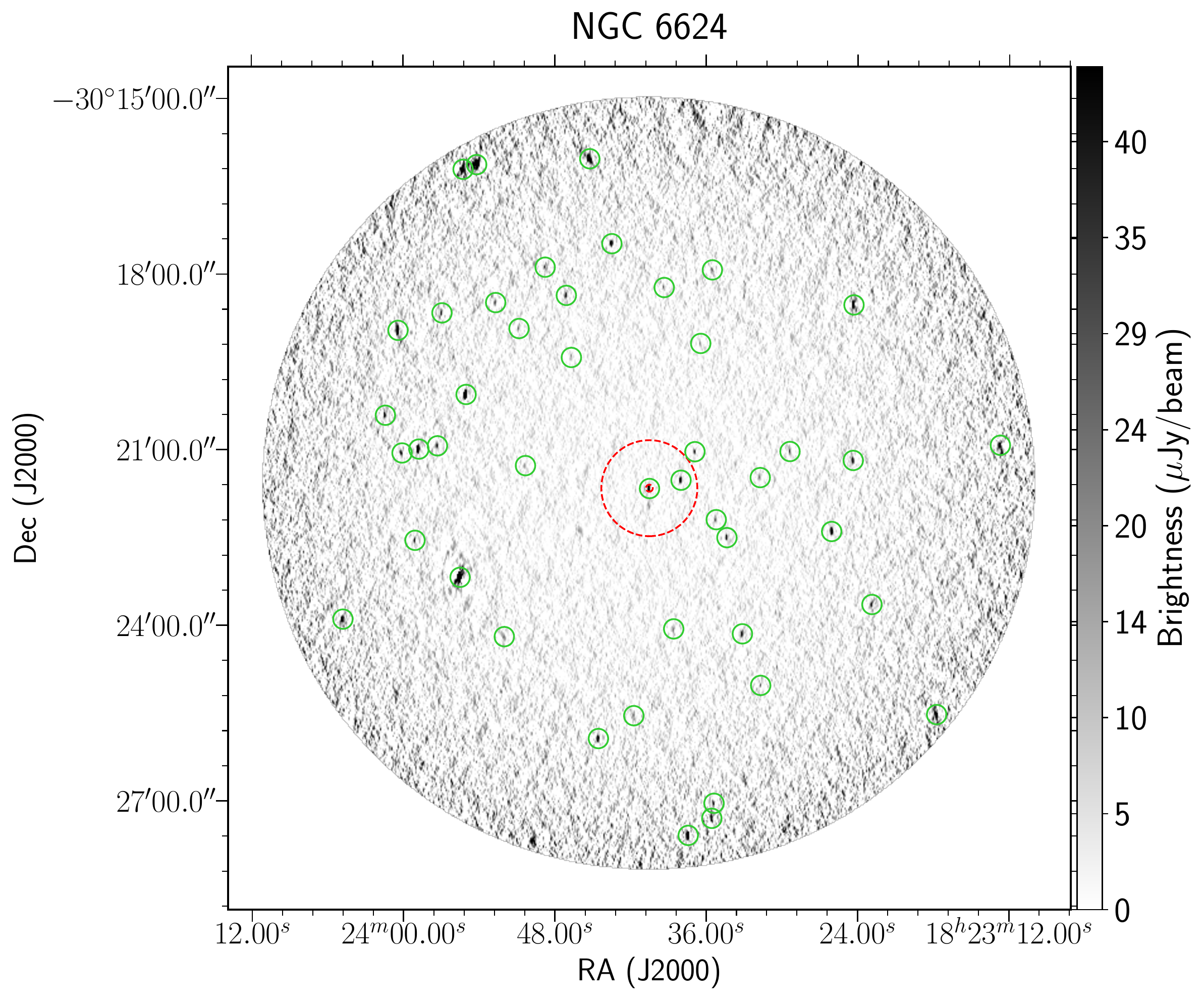}
\end{subfigure}   
\begin{subfigure}{\hsize}
\centering
\includegraphics[width=8.5cm]{figures/appendix_figs/ngc6652_5p5.pdf}
\includegraphics[width=8.5cm]{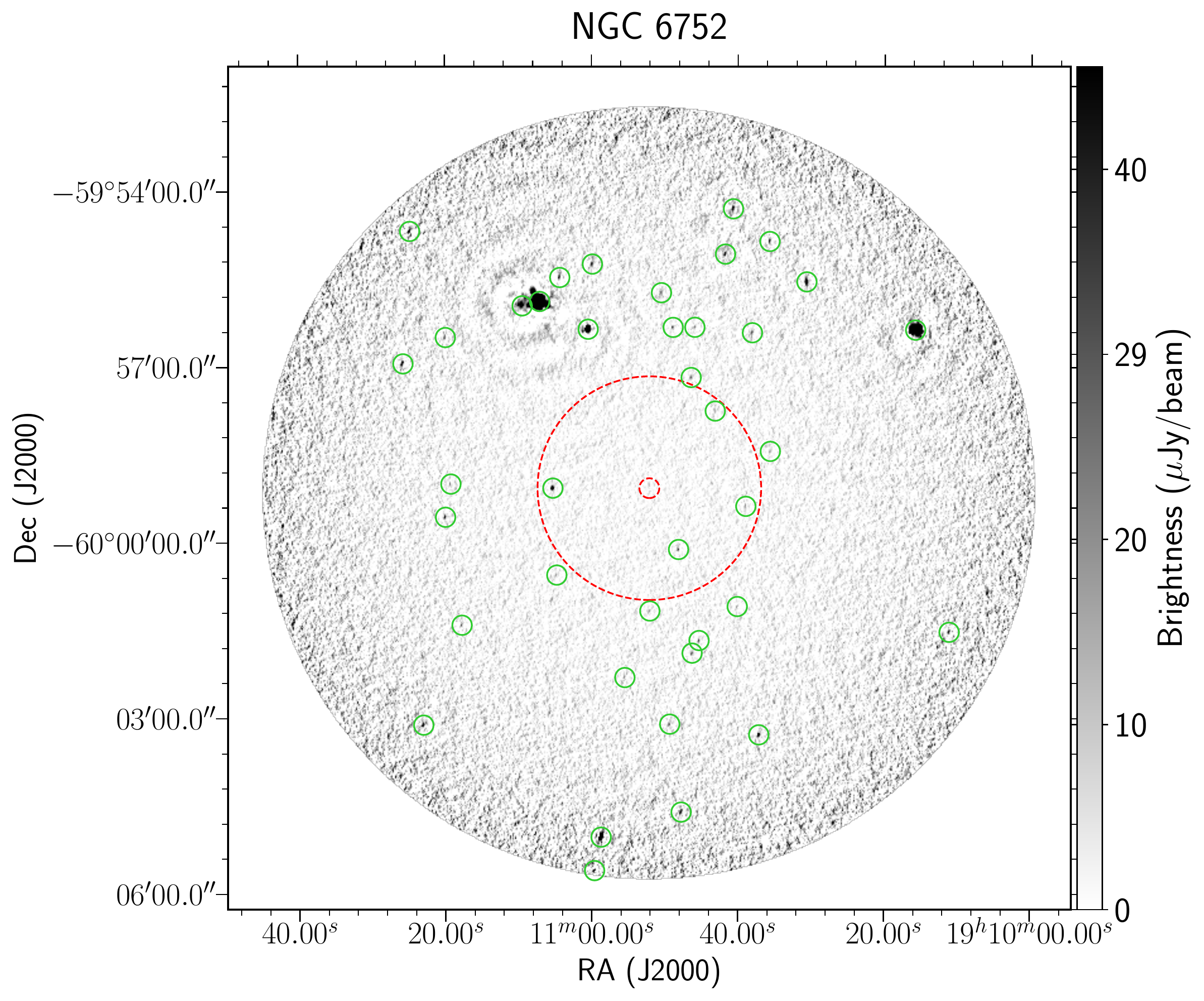}
\end{subfigure}   
\begin{subfigure}{\hsize}
\centering
\includegraphics[width=8.5cm]{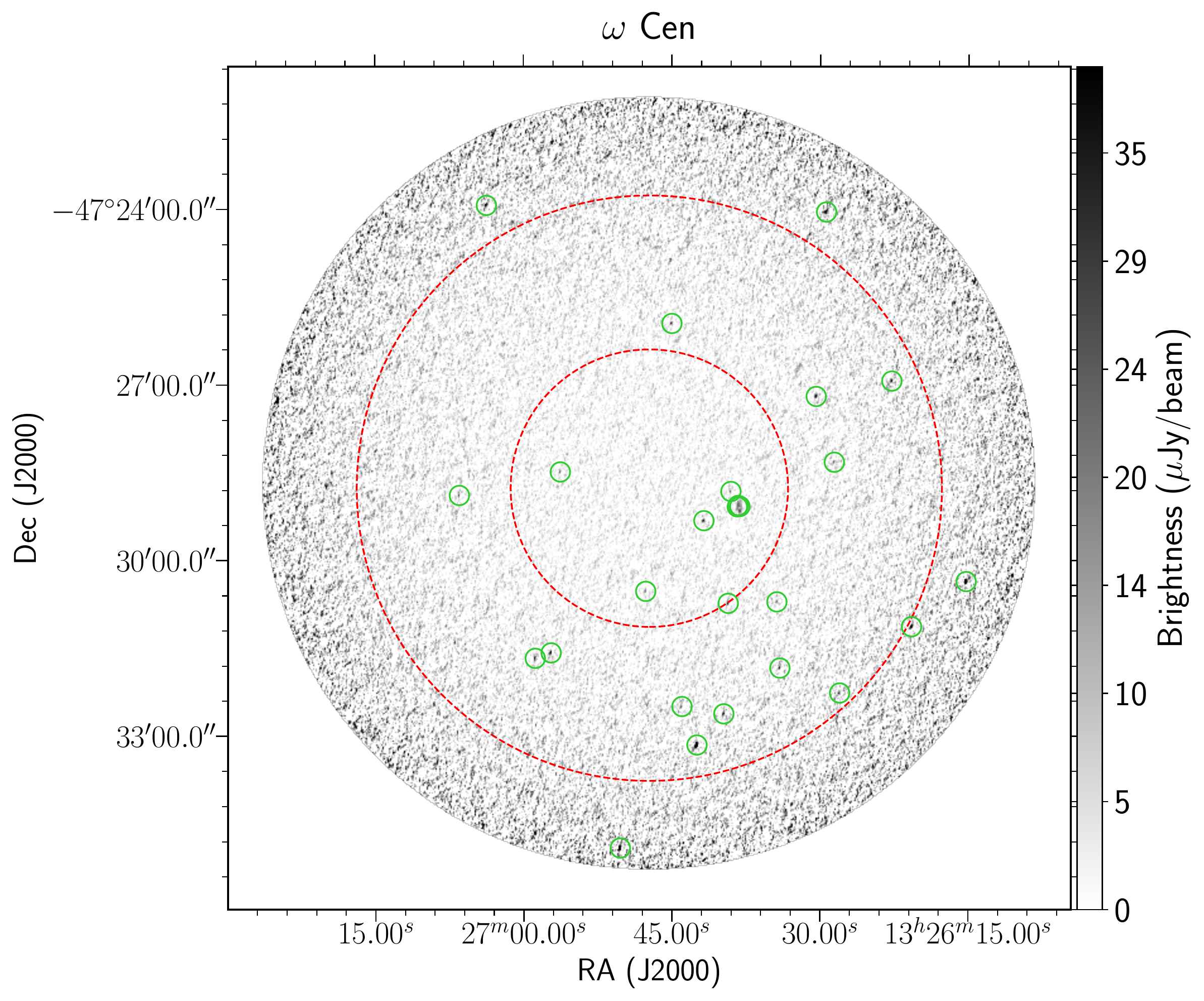}
\includegraphics[width=8.5cm]{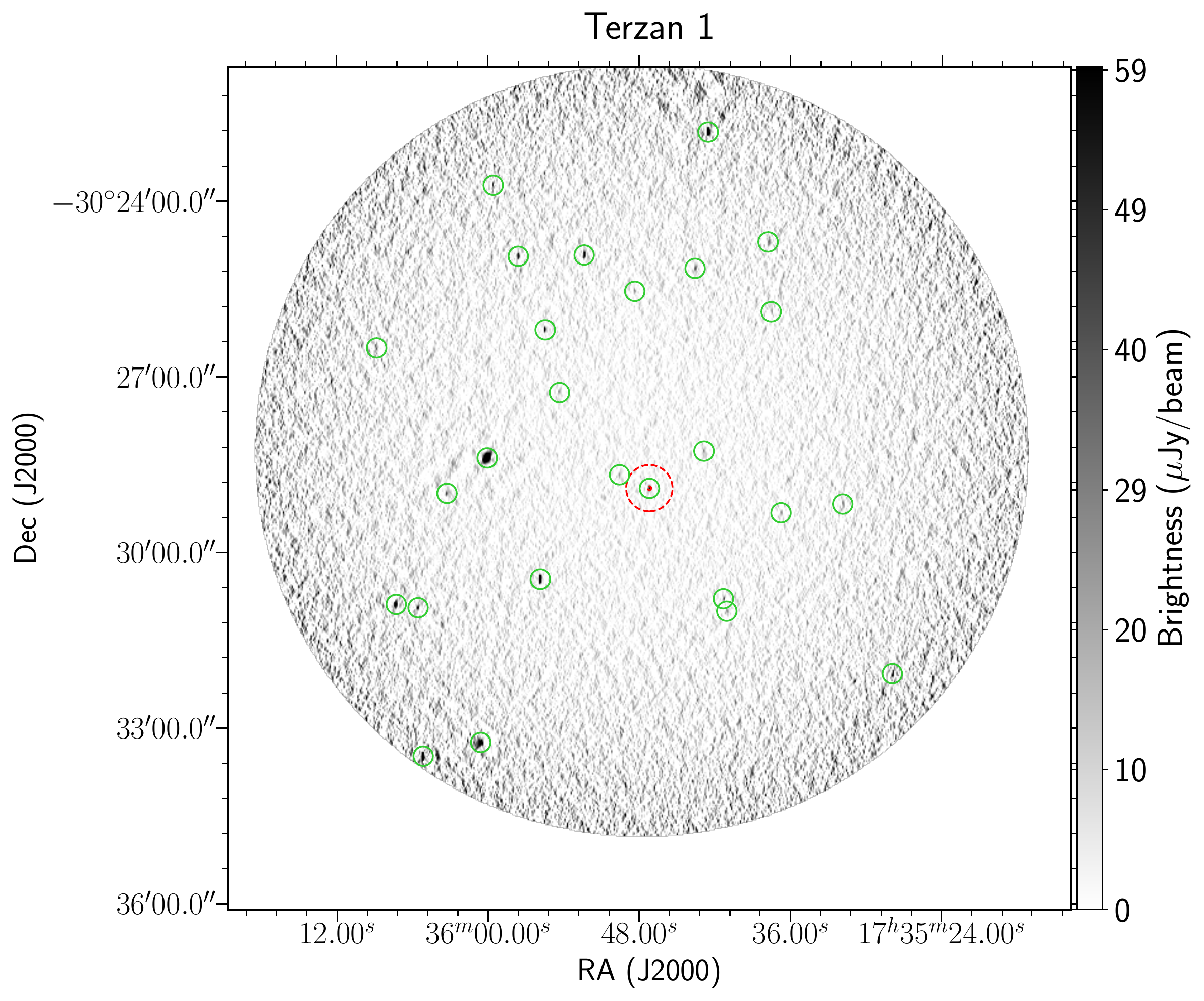}
\end{subfigure}   
\caption{5.5-GHz ATCA images of the fields of NGC\,6553, NGC\,6624, NGC\,6652, NGC\,6752, $\omega$ Cen, and Terzan 1. The core and half-light radii for each cluster are shown by the inner and outer dashed red lines, respectively. Detected sources are highlighted with small green circles. Image brightness is indicated by the colourbar, scaled appropriately for each figure.}
\end{figure*}

\begin{figure*}\ContinuedFloat
\begin{subfigure}[t]{\hsize}
\centering
\includegraphics[width=8.5cm]{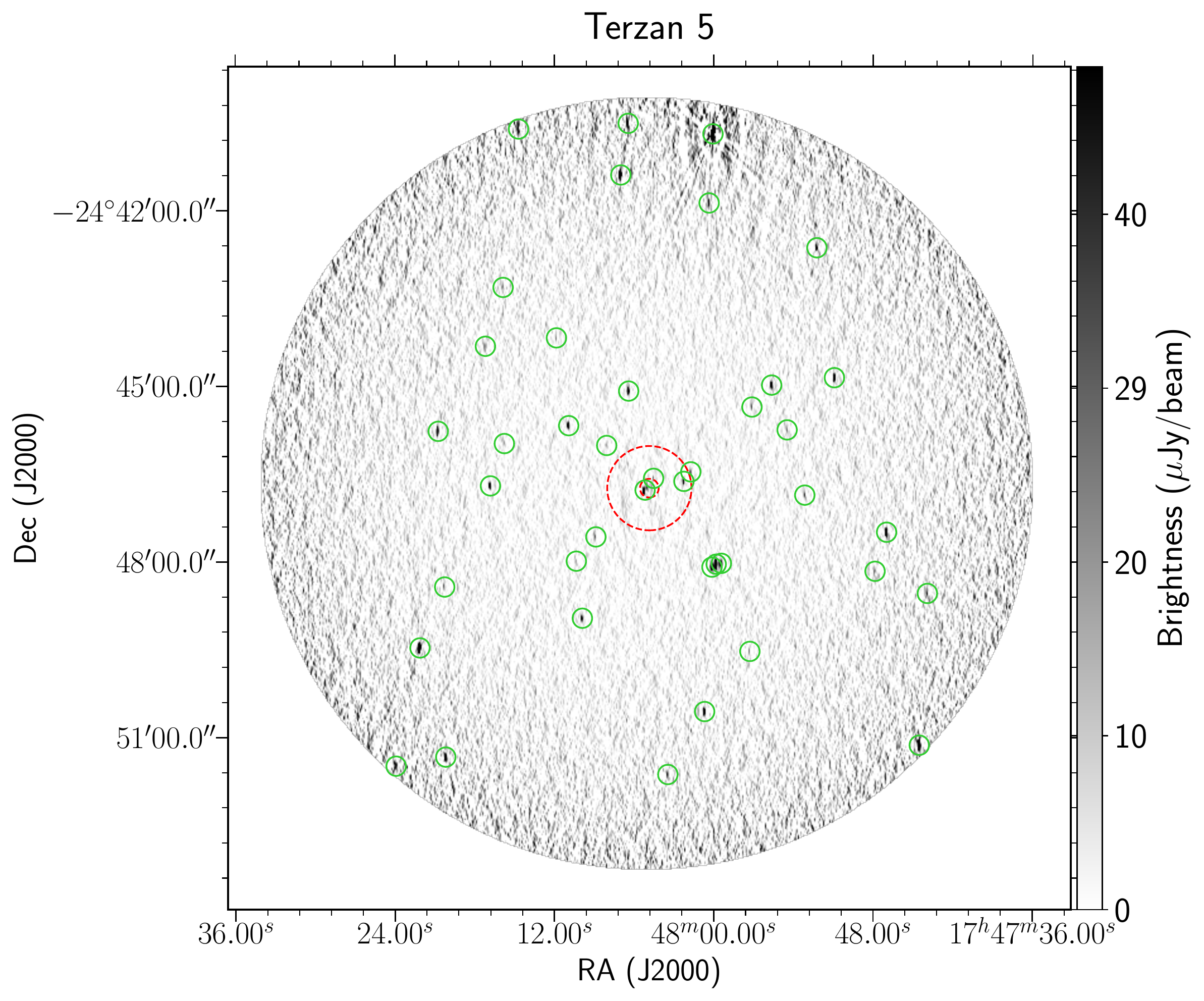}
\includegraphics[width=8.5cm]{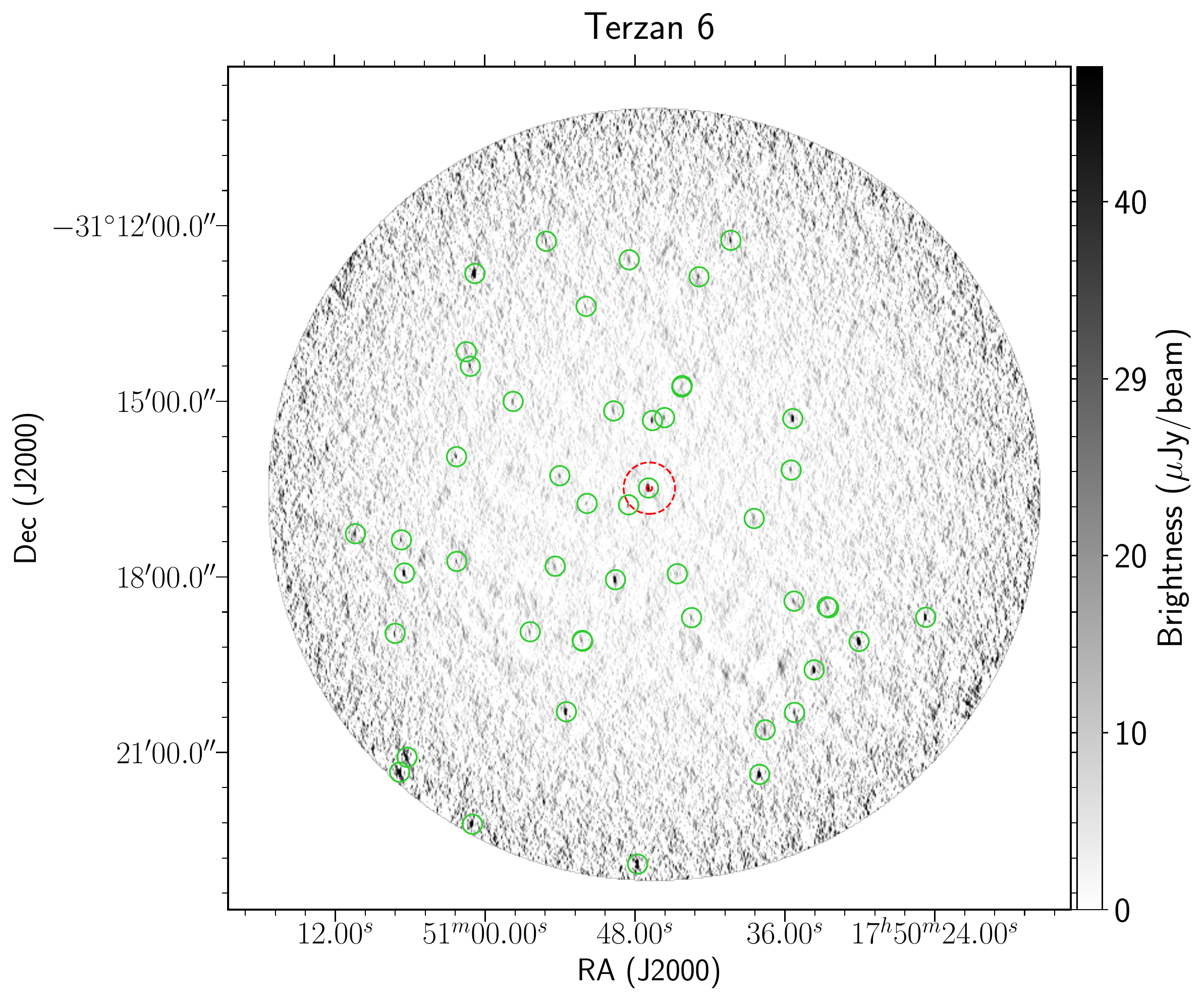}
\end{subfigure}   
\caption{5.5-GHz ATCA images of the fields of Terzan 5 and Terzan 6. The core and half-light radii for each cluster are shown by the inner and outer dashed red lines, respectively. Detected sources are highlighted with small green circles. Image brightness is indicated by the colourbar, scaled appropriately for each figure.}
\label{fig:allradio}
\end{figure*}


\bsp	
\label{lastpage}
\end{document}

%% file: 47Tuc_cat.tex
\begin{tabular}{lllllllllllllrr}
\hline
 ID & 
 \shortstack{RA\\(deg)} & 
 \shortstack{DEC\\(deg)} &  
 \shortstack{$\Delta$\,RA $^{\rm a}$\\({\arcsec})} & \shortstack{$\Delta$\,DEC $^{\rm a}$\\({\arcsec})} & \shortstack{S$_{5.5}$\\(\si{\micro}Jy)} & \shortstack{$\Delta$\,S$_{5.5}$\\(\si{\micro}Jy)} & \shortstack{S$_{7.25}$\\(\si{\micro}Jy)} & \shortstack{$\Delta$\,S$_{7.25}$\\(\si{\micro}Jy)} & \shortstack{S$_{9}$\\(\si{\micro}Jy)} & \shortstack{$\Delta$\,S$_{9}$\\(\si{\micro}Jy)} & $\alpha$ & $\Delta \alpha$ &  R$_{\rm c}$ &  R$_{\rm h}$ \\
\hline\hline
47Tuc-ATCA1 & 6.0180285 & --72.0827214 & 0.24 & 0.04 &  34.7 & 3.8 &  32.6 & 3.2 &  29.1 & 4.7 &  --0.37 & 0.41 & 0.39 & 0.04 \\
47Tuc-ATCA2 & 6.0579037 & --72.0788108 & 0.48 & 0.15 &  24.1 & 3.9 &  16.0 & 3.3 & <17.6 & --- & <-0.64 & --- & 1.79 & 0.20 \\
47Tuc-ATCA3 & 5.9681262 & --72.0656876 & 0.03 & 0.02 &  31.0 & 3.9 &  26.5 & 3.6 & <29.9 & --- & <-0.07 & --- & 3.85 & 0.44 \\
47Tuc-ATCA4 & 6.0556079 & --72.0590536 & 0.28 & 0.07 &  87.7 & 3.9 &  61.8 & 3.7 &  36.3 & 5.9 &  --1.79 & 0.34 & 4.03 & 0.46 \\
47Tuc-ATCA5 & 6.0926497 & --72.0690308 & 0.27 & 0.13 &  24.7 & 4.0 &  21.5 & 3.7 & <33.4 & --- & <0.61 & --- & 4.07 & 0.46 \\
47Tuc-ATCA6 & 6.0755228 & --72.1053155 & 0.47 & 0.09 &  24.6 & 4.2 &  24.7 & 4.0 & <29.6 & --- & <0.38 & --- & 4.82 & 0.55 \\
47Tuc-ATCA7 & 5.9202758 & --72.0865673 & 0.06 & 0.02 &  170 & 4.1 &  120 & 4.2 &  70.7 & 6.8 &  --1.78 & 0.20 & 5.38 & 0.61 \\
47Tuc-ATCA8 & 6.1166201 & --72.1010934 & 0.49 & 0.06 &  29.4 & 4.3 &  26.3 & 4.4 & <36 & --- & <0.41 & --- & 5.80 & 0.66 \\
47Tuc-ATCA9 & 6.0333406 & --72.1180916 & 0.01 & 0.06 & <18.1 & --- &  22.2 & 4.4 &  38.8 & 7.1 & >1.55 & --- & 6.17 & 0.70 \\
47Tuc-ATCA10 & 5.9922285 & --72.1167805 & 0.48 & 0.12 & <16.6 & --- &  18.2 & 4.4 &  40.6 & 7.2 & > 1.82 & --- & 6.15 & 0.70 \\
47Tuc-ATCA11 & 6.1460931 & --72.0837495 & 0.06 & 0.10 &  30.4 & 4.4 &  21.0 & 4.5 & <32.5 & --- & <0.14 & --- & 6.29 & 0.71 \\
47Tuc-ATCA12 & 6.0358483 & --72.1190226 & 0.01 & 0.09 &  28.5 & 4.5 &  22.4 & 4.5 & <25.2 & --- & <--0.25 & --- & 6.34 & 0.72 \\
47Tuc-ATCA13 & 6.0138886 & --72.0429991 & 0.30 & 0.05 &  24.5 & 4.4 &  19.0 & 4.5 & <34.7 & --- & <0.71 & --- & 6.38 & 0.72 \\
47Tuc-ATCA14 & 6.0330011 & --72.0433269 & 0.10 & 0.05 & <18.6 & --- &  18.1 & 4.5 &  41.6 & 7.4 & >1.64 & --- & 6.32 & 0.72 \\
47Tuc-ATCA15 & 6.0553603 & --72.0403011 & 0.16 & 0.09 &  25.0 & 4.5 &  24.8 & 4.7 & <41.3 & --- & <1.02 & --- & 7.00 & 0.80 \\
47Tuc-ATCA16 & 5.9422227 & --72.1208722 & 0.32 & 0.02 &  25.2 & 4.7 &  25.0 & 5.2 & <45.1 & --- & <1.18 & --- & 7.83 & 0.89 \\
47Tuc-ATCA17 & 6.0430120 & --72.1281478 & 0.11 & 0.10 &  71.9 & 4.7 &  58.9 & 5.3 &  50.0 & 9.2 & --0.73 & 0.39 & 7.89 & 0.90 \\
47Tuc-ATCA18 & 5.8689766 & --72.0649243 & 0.43 & 0.10 &  28.2 & 5.0 &  30.6 & 5.5 & <43.6 & --- & <0.89 & --- & 8.39 & 0.95 \\
47Tuc-ATCA19 & 6.1488747 & --72.1194801 & 0.18 & 0.11 &  29.3 & 5.3 &  26.1 & 6.2 & <51.5 & --- & <1.15 & --- & 9.05 & 1.03 \\
47Tuc-ATCA20 & 6.2038151 & --72.0721862 & 0.44 & 0.36 &  31.1 & 5.1 &  21.5 & 6.1 & <47.8 & --- & <0.87 & --- & 9.35 & 1.06 \\
\hline
\end{tabular}

%% file: maveric.bbl
 \newcommand{\noop}[1]{}